\documentclass[longauth]{aa}  

\usepackage{graphicx}
\usepackage{multirow}
\usepackage{tablefootnote}
\usepackage{comment}
\usepackage{gensymb}
 
\usepackage{subcaption}
\usepackage{placeins}

\usepackage{txfonts}
\usepackage{booktabs}
\usepackage{starfont}
\usepackage{soul}

\usepackage[colorlinks=true,citecolor=blue]{hyperref}
\usepackage{orcidlink}

\newcommand{\plaa}{TOI\mbox{-}1243\,b}
\newcommand{\plab}{TOI\mbox{-}4529\,b}
\newcommand{\plac}{TOI\mbox{-}5388\,b}

\newcommand{\splaa}{TOI\mbox{-}1243}
\newcommand{\splab}{TOI\mbox{-}4529}
\newcommand{\splac}{TOI\mbox{-}5388}

\newcommand{\tess}{\textit{TESS}}

\newcommand{\kms}{\mbox{km\,s$^{-1}$}}

\begin{document} 

   \title{Characterization of two new transiting sub-Neptunes and a terrestrial planet around M-dwarf hosts}

   \author{E.~Poultourtzidis\inst{\ref{iac},\ref{iauth}}\thanks{\texttt {epoultou@physics.auth.gr}}\orcidlink{0009-0004-5523-707X},
        G.~Lacedelli \inst{\ref{iac}} \orcidlink{0000-0002-4197-7374},
        E.~Pall{\'e}\inst{\ref{iac}, \ref{ull}} \orcidlink{0000-0003-0987-1593},
        I.~Carleo \inst{\ref{i:torino}, \ref{iac}} \orcidlink{0000-0002-0810-3747},
        C.~Magliano \inst{\ref{i:napoliDF}, \ref{i:napoliU}, \ref{i:napoliObs}} \orcidlink{0000-0001-6343-4744},
        S.~Gerald\'ia-González \inst{\ref{iac}, \ref{ull}} \orcidlink{0009-0002-5545-3034},
        J.~A.~Caballero \inst{\ref{i:csic_inta}} \orcidlink{0000-0002-7349-1387},
        G.~Morello \inst{\ref{iaa}, \ref{i:palermoObs}} \orcidlink{0000-0002-4262-5661},
        J.~Orell-Miquel \inst{\ref{i:texasU}} \orcidlink{0000-0003-2066-8959},
        H.~M.~Tabernero \inst{\ref{i:IEEC}, \ref{i:ice_csic}} \orcidlink{0000-0002-8087-4298},
        F.~Murgas\inst{\ref{iac}, \ref{ull}}  \orcidlink{0000-0001-9087-1245}, 
        G.~Covone \inst{\ref{i:napoliDF}, \ref{i:napoliU}, \ref{i:napoliObs}}  \orcidlink{0000-0002-2553-096X},
        F.~J.~Pozuelos \inst{\ref{iaa}} \orcidlink{0000-0003-1572-7707},
        P.~J.~Amado\inst{\ref{iaa}}\orcidlink{0000-0002-8388-6040},
        V.~J.~S.~B\'ejar\inst{\ref{iac},\ref{ull}} \orcidlink{0000-0002-5086-4232},
        S.~Chairetas  \inst{\ref{iauth}}\orcidlink{0009-0008-5560-6911},
        C.~Cifuentes \inst{\ref{i:csic_inta}}\orcidlink{0000-0003-1715-5087},
        D.~R.~Ciardi \inst{\ref{i:usa_caltech}} \orcidlink{0000-0002-5741-3047},
        K.~A.~Collins \inst{\ref{i:harvardCA}} \orcidlink{0000-0001-6588-9574},
        I.~J.~M.~Crossfield \inst{\ref{i:usa_kansas}},
        E.~Esparza-Borges \inst{\ref{iac}, \ref{ull}} \orcidlink{0000-0002-2341-3233},
        G.~Fernández-Rodríguez \inst{\ref{iac}, \ref{ull}} \orcidlink{0000-0003-0597-7809},
        A.~Fukui \inst{\ref{i:japanKIS}, \ref{iac}} \orcidlink{0000-0002-4909-5763},
        Y.~Hayashi \inst{\ref{i:japanMDS}} \orcidlink{0000-0001-8877-0242},
        A.~P.~Hatzes \inst{\ref{i:germany_TLT}} \orcidlink{0000-0002-3404-8358},
        Th.~Henning\inst{\ref{i:germany_mpia}},
        E.~Herrero \inst{\ref{i:IEEC}} \orcidlink{0000-0001-8602-6639},
        K.~Horne \inst{\ref{i:UK_SUPA}} \orcidlink{0000-0003-1728-0304},
        S.~B.~Howell \inst{\ref{i:usa_ames}}\orcidlink{0000-0002-2532-2853},
        K.~Isogai \inst{\ref{i:japanOO}, \ref{i:japanMDS}} \orcidlink{0000-0002-6480-3799},
        J.~M.~Jenkins\inst{\ref{i:usa_ames}} \orcidlink{0000-0002-4715-9460},
        Y.~Kawai \inst{\ref{i:japanMDS}} \orcidlink{0000-0002-0488-6297},
        F.~Libotte\inst{\ref{iac}, \ref{i:sabadell}},
        E.~Matthews \inst{\ref{i:germany_mpia}}\orcidlink{0000-0003-0593-1560},
        P.~Meni-Gallardo \inst{\ref{iac}, \ref{ull}} \orcidlink{0009-0001-7943-0075}
        I.~Mireles \inst{\ref{i:usa_newMexico}} \orcidlink{0000-0002-4510-2268},
        J.~C.~Morales\inst{\ref{i:IEEC}, \ref{i:ice_csic}}\orcidlink{0000-0003-0061-518X},
        N.~Narita \inst{\ref{i:japanKIS}, \ref{i:japanAC}, \ref{iac}} \orcidlink{0000-0001-8511-2981},
        B.~B.~Ogunwale \inst{\ref{i:uniariel}}\orcidlink{0009-0001-7046-0446},
        H.~Parviainen \inst{\ref{ull}, \ref{iac}} \orcidlink{0000-0001-5519-1391},
        A.~Quirrenbach \inst{\ref{i:germany_LZA}},
        A.~Reiners \inst{\ref{i:germany_IAG}} \orcidlink{0000-0003-1242-5922},
        I.~Ribas\inst{\ref{i:ice_csic}, \ref{i:IEEC}}\orcidlink{0000-0002-6689-0312},
        R.~Sefako \inst{\ref{i:safrica_AO}}\orcidlink{0000-0003-3904-6754},
        A.~Shporer \inst{\ref{i:usa_mit}} \orcidlink{0000-0002-1836-3120}, 
        R.~P.~Schwarz \inst{\ref{i:harvardCA}} \orcidlink{0000-0001-8227-1020},
        G.~Srdoc \inst{\ref{i:croatia_O}},
        L.~Tal-Or\inst{\ref{i:uniariel},\ref{i:astrogeoariel}} \orcidlink{0000-0003-3757-1440},  
        S.~Vanaverbeke \inst{\ref{i:belgiumA},\ref{i:belgiumAstrolab}, \ref{i:belguim_MPA}},
        N.~Watanabe \inst{\ref{i:japanMDS}}\orcidlink{0000-0002-7522-8195}, 
        C.~N.~Watkins \inst{\ref{i:harvardCA}} \orcidlink{0000-0001-8621-6731},
        F.~Zong~Lang\inst{\ref{i:bern_habitat}}
            }

    \institute{Instituto de Astrof\'{i}sica de Canarias (IAC), 38205 La Laguna, Tenerife, Spain\label{iac} \email{epoultou@physics.auth.gr}
        \and Departamento de Astrof\'isica, Universidad de La Laguna (ULL), 38206 La Laguna, Tenerife, Spain\label{ull}
        \and Department of Physics, Aristotle University of Thessaloniki, University Campus, Thessaloniki, 54124, Greece\label{iauth}
        \and  INAF - Osservatorio Astrofisico di Torino, Via Osservatorio 20, 10025 Pino Torinese, Italy\label{i:torino}
        \and Dipartimento di Fisica ``Ettore Pancini'', Università di Napoli Federico II, Napoli, Italy\label{i:napoliDF}
        \and INFN, Sezione di Napoli, Complesso Universitario di Monte S. Angelo, Via Cintia Edificio 6, 80126, Napoli, Italy \label{i:napoliU}
        \and INAF - Osservatorio Astronomico di Capodimonte, via Moiariello 16, 80131, Napoli, Italy \label{i:napoliObs}
        \and Centro de Astrobiolog\'ia (CSIC-INTA), European Space Astronomy Centre, Camino Bajo del Castillo, 28692 Villanueva de la Ca\~nada, Madrid, Spain 
        \label{i:csic_inta}
        \and Instituto de Astrof\'isica de Andaluc\'ia (IAA), 18080, Granada, Spain\label{iaa}
        \and INAF, Osservatorio Astronomico di Palermo, 90134 Palermo, Italy \label{i:palermoObs}
        \and Department of Astronomy, University of Texas at Austin, 2515 Speedway, Austin, TX 78712, USA \label{i:texasU}
        \and Institut d'Estudis Espacials de Catalunya (IEEC), C/ Esteve Terrades 1, Edifici RDIT, 08860 Castelldefels, Spain \label{i:IEEC}
        \and Institut de Ci\`encies de l’Espai (ICE, CSIC), Campus UAB, c/ de Can Magrans s/n, 08193 Cerdanyola del Vall\`es, Barcelona, Spain \label{i:ice_csic}
        \and NASA Exoplanet Science Institute – Caltech/IPAC 1200 E. California Blvd Pasadena, CA 91125, USA \label{i:usa_caltech}
        \and Center for Astrophysics | Harvard \& Smithsonian, 60 Garden Street, Cambridge, MA 02138, USA \label{i:harvardCA}
        \and Department of Physics and Astronomy, University of Kansas, Lawrence, KS, USA\label{i:usa_kansas}
        \and Vereniging Voor Sterrenkunde, Oude Bleken 12, 2400 Mol, Belgium\label{i:belgiumA}
        \and AstroLAB IRIS, Provinciaal Domein ``De Palingbeek'', Verbrande-molenstraat 5, 8902 Zillebeke, Ieper, Belgium \label{i:belgiumAstrolab}
        \and Komaba Institute for Science, The University of Tokyo, 3-8-1 Komaba, Meguro, Tokyo 153-8902, Japan \label{i:japanKIS}
        \and Department of Physics and Kavli Institute for Astrophysics and Space Research, Massachusetts Institute of Technology, Cambridge, MA 02139, USA \label{i:usa_mit}
        \and Department of Multi-Disciplinary Sciences, Graduate School of Arts and Sciences, The University of Tokyo, 3-8-1 Komaba, Meguro, Tokyo 153-8902, Japan \label{i:japanMDS}
        \and Th\"uringer Landessternwarte Tautenburg, Sternwarte 5, 07775 Tautenburg, Germany \label{i:germany_TLT}
        \and Max-Planck-Institut f\"ur Astronomie, K\"onigstuhl 17, 69117 Heidelberg, Germany \label{i:germany_mpia}
        \and SUPA Physics and Astronomy, University of St. Andrews, Fife, KY16 9SS, Scotland, UK \label{i:UK_SUPA}
        \and NASA Ames Research Center, Moffett Field, CA 94035, USA \label{i:usa_ames}
        \and Okayama Observatory, Kyoto University, 3037-5 Honjo, Kamogatacho, Asakuchi, Okayama 719-0232, Japan \label{i:japanOO}
        \and Agrupaci\'o Astronomica Sabadell, Carrer Prat de la Riba, 116, E-08206 Sabadell, Barcelona, Spain\label{i:sabadell}
        \and Department of Physics and Astronomy, The University of New Mexico, 210 Yale Blvd. NE., Albuquerque, NM 87106, USA \label{i:usa_newMexico} 
        \and Astrobiology Center, 2-21-1 Osawa, Mitaka, Tokyo 181-8588, Japan\label{i:japanAC}
        \and Department of Physics, Ariel University, Ariel 40700, Israel \label{i:uniariel} 
        \and Landessternwarte, Zentrum f\"ur Astronomie der Universit\"at Heidelberg, 69117 Heidelberg, Germany \label{i:germany_LZA} 
        \and Institut f\"ur Astrophysik und Geophysik Georg-August-Universit\"at, Friedrich Hund Platz 1,l 37077 G\"ottingen, Germany \label{i:germany_IAG}
        \and South African Astronomical Observatory, P.O. Box 9, Observatory, Cape Town 7935, South Africa \label{i:safrica_AO}
        \and Kotizarovci Observatory, Sarsoni 90, 51216 Viskovo, Croatia \label{i:croatia_O} 
        \and Astrophysics, Geophysics, And Space Science Research Center, Ariel University, Ariel 40700, Israel \label{i:astrogeoariel}
        \and Centre for Mathematical Plasma-Astrophysics, Department of Mathematics, KU Leuven, Celestijnenlaan 200B, 3001 Heverlee, Belgium \label{i:belguim_MPA}
        \and Center for Space and Habitability, University of Bern, Gesellschaftsstrasse 6, 3012, Bern, Switzerland\label{i:bern_habitat}
        }

   \date{Received 08 September 2025 / Accepted dd Month 2025}

    \abstract 
  {We report the confirmation of three transiting exoplanets orbiting \splaa\ (LSPM~J0902+7138), \splab\ (G~2--21), and \splac\ (Wolf~346) that were initially detected by \tess through ground-based photometry and radial velocity follow-up measurements with CARMENES. The planets present short orbital periods of $4.65$, $5.88$, and $2.59$ days, and they orbit early-M dwarfs (M2.0\,V, M1.5\,V, and M3.0\,V, respectively). We were able to precisely determine the radius of all three planets with a precision of $< 7\, \%$, the mass of \plaa\ with a precision of $19\, \%$, and upper mass limits for \plab\ and \plac. The radius of \plaa\ is $2.33\pm0.12$ \, $\mathrm{R_{\oplus}}$, its mass is $7.7 \pm 1.5\,\mathrm{M_{\oplus}}$, and the mean density is $0.61 \pm 0.15$ \, $\mathrm{\rho_\oplus}$. The radius of \plab\ is $1.77 ^{+0.09}_{-0.08} \, \mathrm{R_{\oplus}}$, the $3 \mathrm{\sigma}$ upper mass limit is $4.9 \, \mathrm{M_{\oplus}}$, and the $3 \mathrm{\sigma}$ upper density limit is $0.88\, \mathrm{\rho_\oplus}$. The third planet, \plac, is Earth-sized with a radius of $0.99 ^{+0.07}_{-0.06} \, \mathrm{R_{\oplus}}$, a $3 \mathrm{\sigma}$ upper mass limit of $2.2 \, \mathrm{M_{\oplus}}$ ,and a $3 \mathrm{\sigma}$ upper density limit of $2.2\, \mathrm{\rho_\oplus}$. While \plac\ is most probably rocky, given its Earth-like radius, \plaa\ and \plab\ are located in a highly degenerate region in the mass-radius space. \plab\ appears to lean toward a water-world composition. \plaa\ has enough mass to host a significant H-He envelope, although a water-world and pure rocky compositions are also consistent with the data. Our analysis indicates that future atmospheric observations using JWST can aid in determining their real composition. The sample of small planets around M dwarfs is widely used to understand planet formation and composition theories, and our study adds three planets to this sample.
}

   \keywords{planets and satellites: detection -- techniques: photometric -- techniques: radial velocities -- stars: individual: TOI-1243, TOI-4529, TOI-5388 
               }
\titlerunning{TOI-1243, TOI-4529, TOI-5388}
\authorrunning{Poultourtzidis et al.}
   \maketitle

\section{Introduction}

The discovery of a large population of well-characterized exoplanets has significantly advanced our understanding of planetary formation and evolution processes. Observational data acquired in the past decades, mainly from space missions such as \textit{Kepler} \citep{kepler2010} and the Transiting Exoplanet Survey Satellite (\tess;  \citealt{tess_2015}), highlighted a population of exoplanets with radii between those of Earth and Neptune ($1\,{\rm R}_\oplus < R_{p} < 4\,{\rm R}_\oplus$). They are called sub-Neptunes or super-Earths. This category of planets is absent from our Solar System. More than half of the Sun-like stars in the Galaxy host a sub-Neptune within orbits closer than 1 \,au, however \citep{bathala2013, Petigura_2013, marcy2014}. For this reason, a precise characterization of a substantial population of sub-Neptunes is crucial for better understanding their nature and for providing observational evidence to test proposed or emerging theories of planetary system formation and evolution.

Recent demographic studies revealed a paucity of sub-Neptunes with radii between 1.5\,R$_\oplus$ and 2.0\,R$_\oplus$, which is known as the radius gap (\citealt{fulton_2017, val_eulen_2018, hirano_2018, berger2018}). Various hypotheses attempted to explain this observational phenomenon by assuming a loss of primordial atmosphere (e.g., \citealt{owen2013, lopez2013, ginzburg2013, gupta2013}), a different internal composition (e.g., \citealt{mousis2020, dorn2021, Izidoro_2022}), or accretion mechanisms \citep{lee_2022_radiusGap}.

Using a revised sample of small exoplanets ($R_{\rm p} < 4\,{\rm R}_\oplus$), \cite{water_worlds_2022} recently claimed that the radius gap around M-dwarf stars might be a consequence of different interior compositions and not evidence of atmospheric loss. This study divided the planets into three main categories: (1) rocky planets, with compositions similar to that of Earth, (2) planets composed of equal quantities of silicates and water, called water worlds, and (3) mini-Neptunes with significant hydrogen-helium envelopes.   
Using an updated sample, \cite{parc2024} claimed that water worlds around M dwarfs cannot correspond to a distinct population because their bulk density and equilibrium temperature can be interpreted by several different internal structures and compositions, as was also shown by \cite{rogers_2023}, who was able to explain the sub-Neptunes distribution of the \cite{water_worlds_2022} sample using mass-loss theories.

These studies showed that sub-Neptunes around M-dwarf stars are at the epicenter of understanding the internal composition and evolution of planets and planetary systems. It is therefore crucial to increase the sample of precisely characterized sub-Neptunes around M dwarfs to help us distinguish between the different theories. In this context, the synergies between \tess\ and ground-based high-precision infrared spectrographs such as the Calar Alto high-Resolution search for M dwarfs with Exoearths with Near-infrared and optical Echelle Spectrographs (CARMENES; \citealt{carmenes_2010, carmeness_2014}) have provided a large number of small exoplanet discoveries around M-dwarf stars, which are also golden targets for atmospheric studies \citep{carm_luq_2019, carm_luque2022, carm_nowak_2020, Trifonov2021, carm_gonzalez_2022, carm_palle2023}. 
In the era of the \textit{James Webb} Space Telescope (\textit{JWST}; \citealt{jwst_2006}), the importance of observing the atmospheres of exoplanets around M dwarfs has already been highlighted (e.g., \citealt{Lustig_Yaeger_2023, Moran_2023}). New discoveries and precise ephemerides can increase the current sample suitable for atmospheric characterization and provide new potential targets for current and future instruments tailored for observing exoplanetary atmospheres such as \textit{JWST}, \textit{Ariel} \citep{ariel_2018}, or the ArmazoNes high Dispersion Echelle Spectrograph (ANDES; \citealt{andes2024,Palle2025}).
In this study, we characterize three exoplanets around the M dwarfs \splaa\ (LSPM~J0902+7138), \splab\ (G~2--21), and \splac\ (Wolf~346) using \tess\ (Sect.~\ref{section:tess_photometric_data}), combined with  ground-based photometric data and CARMENES radial velocity (RV) measurements (Sect.~\ref{section:follow_up_observations}). Our analysis confirms the planetary nature of these candidates (Sect.~\ref{sec:photometric_vetting}), provides precise stellar parameters (Sect.~\ref{sec:star}), and determines precise planetary radii and masses when possible, or upper mass limits (Sect.~\ref{section:data_analysis}). These place the new discoveries in the context of the known sub-Neptune sample (Sect.~\ref{sec:discussion}).

\section{\textit{TESS} photometry}
\label{section:tess_photometric_data}

All three candidates were announced as \tess\ objects of interest (TOI) by the Science Processing Operations Center pipeline (SPOC; \citealt{spoc2016}) using a wavelet-based noise-compensating matched filter \citep{jekins2002, 2020TPSkdph}.
The transit signatures were fit with initial limb-darkened transit models (\citealt{Li:DVmodelFit2019}) and subjected to a suite of diagnostic tests to help us confirm or refute the planetary interpretation of the data (\citealt{Twicken:DVdiagnostics2018}). \splab\ was jointly detected by the \tess\ Quick-Look Pipeline (QLP; \citealt{huang2020}). The candidates passed the data validation tests, including the difference-image analysis tests, which located the host star to within $2.1 \pm 3.4''$ (Sectors 14--74), $1.3 \pm 2.8 ''$ (Sectors 42--70), and $63 \pm 2.9 ''$ (Sector 48) for \plaa, \plab, and \plac, respectively. The \tess\ Science Office reviewed the data validation reports and issued TOI alerts for all three candidates (\citealt{guerrero:TOIs2021ApJS}). Details about the \tess\ observations can be found in Table~\ref{tab:tess_observations}.

We used the 2-minute cadence light curves (LCs) to validate the planets, study the stellar activity, investigate the stellar rotational period, and fit the LC. We used the simple aperture photometry (SAP) and the presearch data conditioning simple aperture photometry (PDCSAP; \citealt{smith2012,Stumpe2012,stumpe2014}) as computed by the SPOC pipeline.

Moreover, we performed an independent search for planetary candidates in the \tess\ data using the \texttt{SHERLOCK}\footnote{\url{https://github.com/franpoz/SHERLOCK}} package \citep{pozuelos2020,devora2024}. For each TOI, we combined all available 2-minute cadence sectors and searched for signals with orbital periods ranging from 0.4 to 30\,days. We explored ten different detrending scenarios with window sizes from 0.2 to 1.2\,days. In all cases, we successfully recovered the TOIs and independently confirmed the alerts. We did not detect any additional signals that might be attributed to new planetary candidates, however.

\section{Ground-based follow-up observations}
\label{section:follow_up_observations}

\subsection{Transit photometry}
\label{ssec:transit_photometry}
We list here the ground-based follow-up photometric transit observations of the three candidates, which we performed as part of the \tess\ follow-up observing program\footnote{\url{https://tess.mit.edu/followup}} \citep[TFOP;][]{collins:2019}. A summary of the ground-based follow-up transit observations can be found in Table \ref{tab:ground_based_transits}, and plots of the detrended transits are shown in Appendix~\ref{appendix:ground_based_lcs}. All ground-based LCs are available at ExoFOP\footnote{\url{https://exofop.ipac.caltech.edu/tess/target.php?id=219698776}}.

\subsubsection{MuSCAT2}
The Multicolor Simultaneous Camera for studying Atmospheres of Transiting exoplanets 2 (MuSCAT2; \citealt{narita2019}) is a four-band $g'$, $r'$, $i'$, $z_s'$ instrument mounted on the 1.52\,m Telescopio Carlos S\'anchez (TCS) at Teide Observatory in Tenerife, Canary Islands, Spain. We observed with MuSCAT2 two full transits of \plab\ on 2022\,January\,21 and 2022\,October\,30, and two full transits of \plac\ on 2023\,March\,13 and 2024\,April\,26.
The data calibration and photometric analysis were performed using the MuSCAT2 photometric pipeline (\citealt{Parviainen2015,Parviainen_2020}).

\subsubsection{LCOGT}

We observed one full transit of \plaa\ on 2021\,February\,3 in Pan-STARRS $i'$ band from the Las Cumbres Observatory Global Telescope \citep[LCOGT;][]{brown2013} 1\,m network node at McDonald Observatory (McD) near Fort Davis, Texas, USA. 
Another full transit window was observed on 2022\,February\,12 simultaneously in Sloan $g'$, $r'$, $i'$ and Pan-STARRS $z$-short ($z_s$) with the MuSCAT3 \citep{muscat3_2020} camera on the 2\,m Faulkes Telescope North at Haleakal\={a} Observatory (HO) on Maui, Hawai'i, USA. We observed a partial and a full transit of \plab\ on 2022\,December\, 17 and 2023\,September\, 13 in Pan-STARRS $z_s$ band from LCOGT 1\,m network nodes at Cerro Tololo Inter-American Observatory in Chile (CTIO) and Siding Spring Observatory near Coonabarabran, Australia (SSO), respectively. Two full transits of \plac\ were observed on 2022\,March\,26 and 2022\,April\,08 in Sloan $i'$ band from LCOGT 1\,m network nodes at McD and CTIO. We observed another full transit on 2022\,April\,15 in Pan-STARRS $z_s$ band from LCOGT 1\,m network node at Observatorio del Teide in Tenerife (TEID). All images were calibrated by the standard LCOGT {\tt BANZAI} pipeline \citep{McCully:2018}, and differential photometric data were extracted using {\tt AstroImageJ} \citep{Collins:2017}. 

\subsubsection{SAINT-EX}

Follow-up transit photometry was performed for \plab\ and \plac, for which we observed one transit each. The observations were taken with the 1\,m SAINT-EX telescope located at the Observatorio Astron\'omico Nacional, in the Sierra de San Pedro M\'artir in Baja California, Mexico. We observed one partial transit for \plab\ in the $z_s'$ band and one full transit for \plac\ in $r'$ band. The data were reduced using the instrument custom pipeline \texttt{PRINCE} \citep{demory2020}. The LCs we used for our analysis were further corrected for systematics using a principal component analysis (PCA) method based on the LCs of all suitable stars in the field of view, except for the target star \citep{wells2021}.

\subsection{Long-term follow-up photometry}
\label{ssec:long_term_photometry}
Long-term photometric data were used to identify and model stellar activity signals mainly caused by stellar rotation, and to correct for them in the planet orbit RV fitting. A summary of all the long-term photometric observations we used can be found in Table~\ref{tab:ground_based_longterm}.

\subsubsection{TJO}

We observed \splaa\ and \splab\ with the 0.8\,m Telescopio Joan Or\'o \citep[TJO,][]{colome2010} at the Observatorio del Montsec in Lleida, Spain, from 2024\,March to 2025\,January, and \splac\ from 2023\,April to 2025\,January. A total of 774 images for \splaa, 343 images for \splab, and 809 images for \splac\ were obtained with the Johnson \textit{R} filter using the LAIA imager. Raw frames were corrected for dark current and bias and were flat-fielded using the \texttt{ICAT} pipeline \citep{colome_ribas2006} of the TJO. The aperture photometry was extracted with \texttt{AstroImageJ} by using an optimal aperture size that minimized the root mean square error (RMS) of the resulting relative fluxes. To derive the differential photometry of the three monitored targets, we selected the 10 to 15 of the brightest comparison stars in each case that did not vary. Then, we removed outliers and measurements affected by poor observing conditions or a low signal-to-noise ratio. The RMS of the TJO differential photometry after the removal of outliers were 14\,mmag, 11\,mmag, and 11\,mmag for \splaa, \splab\, and \splac, respectively.

\subsubsection{e-EYE}

\splaa, \splab, and \splac\ were also observed from the remote telescope hosting facility Entre Encinas y Estrellas (e-EYE) located at Fregenal de la Sierra in Badajoz, Spain\footnote{\url{www.e-eye.es/}}. We used a  $16''$ ODK-corrected Dall-Kirkham reflector and collected $V$- and $R$-band observations with a Kodak KAF-16803 CCD chip mounted on ASA DDM85. The images were reduced and differential photometry was performed using the package {\tt LesvePhotometry}\footnote{\url{www.dppobservatory.net}}. We obtained 147 epochs for \splaa\ from 2024\,March to 2025\,January, 129 epochs for \splab\ from 2024\,May to 2025\,January and 121 epochs for \splac\ from 2024\,March to 2025\,January, respectively.

\subsubsection{LCOGT}

We performed long-term monitoring of \splaa, \splab, and \plac\ with the LCOGT network. \splaa\ was observed during 48 nights between 2024\,March\,22 and 2024\,June\,3 using two telescopes at McD in the $B$ filter with an exposure time of 66\,s. \splab\ was observed during 47 nights between 2024\,September\,22 and 2024\,December\,27 using two telescopes at McD in the $B$ filter with an exposure time of 60\,s. \splac\ was observed during 41 nights between 2024\,March\,26 and 2024\,June\,14 using two telescopes at McD with the $V$ filter and an exposure time of 42\,s.

We also monitored \splaa, \splab, and \splac\ using the 0.4\,m telescopes of the LCOGT network. Observations were conducted in the $V$ band with QHY600 cameras \citep{Harbeck2024}, configured with a $30\arcmin \times 30\arcmin$ field of view. \splaa\ was observed from McD and TEID on 48 occasions between 2024\,April\,4 and 2024\,May\,25, with each visit comprising three exposures of 600\,s. \splab\ was monitored from five LCOGT observatories (CTIO, SSO, McD, TEID, HO, and the South African Astronomical Observatory) for over 93 visits from 2024\,August\,21 to 2024\,December\,27 using five exposures of 150\,s per visit. \splac\ was observed from HO and TEID during 22 visits between 2024\,April\,7 and 2024\,June\,6, with five exposures of 300\,s per visit. All images were calibrated with {\tt BANZAI}, and the differential aperture photometry was then performed with {\tt AstroImageJ}.

\subsection{NIRI high-contrast imaging}
\label{ssec:high_contrast}

\begin{figure}[]
    \centering
        \includegraphics[width=0.43\textwidth]{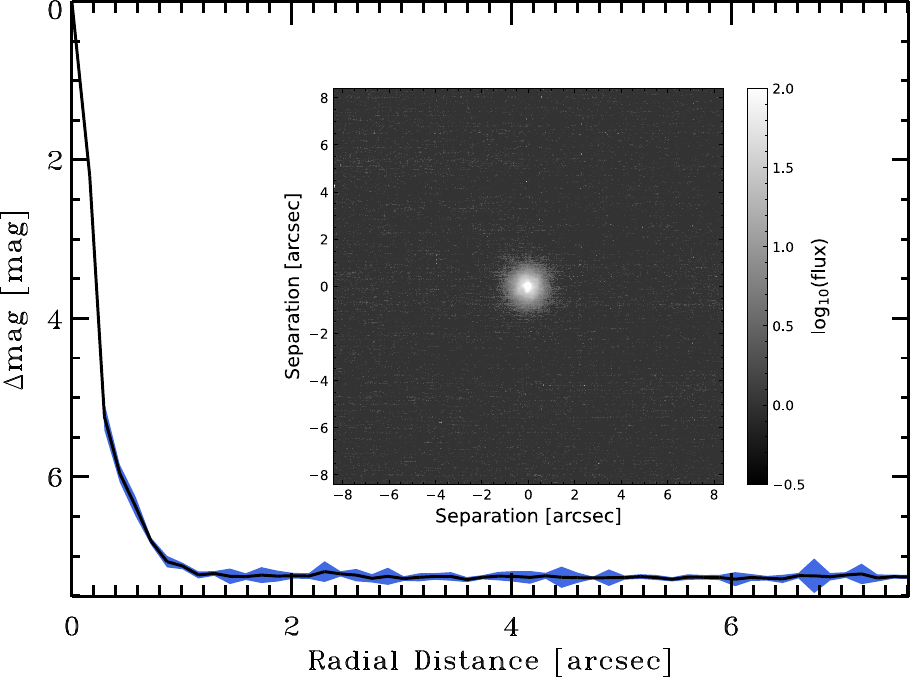}
        \caption{Br$\gamma$ Gemini North NIRI adaptive-optics image and $5 \sigma$ contrast limits for \splaa.}
        \label{fig:hci-1243}
\end{figure}

We observed \splaa\ with the Gemini North NIRI adaptive-optics imager \citep{hodaph2003} on 2019\,November\,25. We collected a total of nine frames in the Br$_\gamma$ filter ($\overline\lambda$ = 2.1686\,$\mu$m), with individual exposure times of 3.6\,s. The telescope was dithered by roughly $3''$ between each frame in a square grid. To calibrate each frame, we applied standard flat-field correction, removed bad pixels, and subtracted the sky background. We then aligned the frames to the position of the star in each image and coadded the nine science frames. The final image reached a contrast floor of roughly 7\,mag at a separation of $0.8''$, and no additional point sources were detected. The raw data are available in the Gemini data archive, and the stacked image and contrast curve (Fig.~\ref{fig:hci-1243}) are available at ExoFOP.

\subsection{CARMENES spectroscopic data}

\splaa, \splab, and \splac\ were observed using the CARMENES (\citealt{carmeness_2014, caballero2025}) spectrograph, installed at the 3.5\,m telescope at the Observatorio de Calar Alto in Almer\'ia, Spain. CARMENES has two spectral channels: an optical channel (VIS) covering wavelengths from $0.52\,\mu \mathrm{m}$ to $0.96\,\mu \mathrm{m}$m with a resolving power of $\mathcal{R} = 94\,600$, and a near-infrared channel (NIR) spanning from $0.96\,\mu \mathrm{m}$ to $1.71\,\mu \mathrm{m}$ with a resolving power of $\mathcal{R} = 80\,400$. Observations have a typical signal-to-noise ratio (S/N) of 41--188 at about 7370\,\AA. More details about the observing campaign for the three planets can be found in Table~\ref{tab:tab:observing_params}. 

The observations were reduced with the CARMENES pipeline \texttt{caracal} \citep{Caballero2016}, and the VIS and NIR spectra were processed with \texttt{serval}\footnote{\url{https://github.com/mzechmeister/serval}} \citep{SERVAL}, which is the standard CARMENES pipeline to derive relative RVs and activity indicators, that is, chromatic radial velocity index (CRX), differential line width (dLW), and H$\alpha$, Na\,D1 \& Na\,D2, and \ion{Ca}{II}\,IRT line indices. \texttt{serval} RVs were further corrected using measured nightly zeropoint corrections, as discussed by \citealt{Trifonov2020}. For the RV fit of the three planets, we employed the data in the VIS channel alone because their internal precision is better than that of the NIR channel (\citealt{bauer2020,ribas2023,caballero2025}) for all three planets. We did use the NIR data for the activity-indicator analysis and to determine the stellar parameters, however. All the RVs and activity indicators are available online at the Centre de Donn\'ees astronomiques de Strasbourg (CDS).

\section{Photometric vetting}
\label{sec:photometric_vetting}

\subsection{Background analysis}

The \tess\ orbital trajectory is designed to optimize sky coverage and minimize stray light. The \tess\ full-frame images might still encounter occasional contamination, however, primarily from zodiacal light and scattered light from Solar System objects \citep{Gangestad2013,Sullivan2015}. Consequently, the background flux can vary during the observation period of each \tess\ sector. To ensure the integrity of the signal, we conducted a visual inspection of a 3-day segment of the background flux centered around the transit time. We examined the background flux for all the transits of the three candidates, but found no spikes or uncertain events that coincided with each transit. Finally, we found no known \tess\ momentum dumps\footnote{A momentum dump is an operational manoeuvrer of a spacecraft that helps reducing the angular velocity of its reaction wheels. More about the \tess\ sector specific momentum dumps can be found in the data release notes for each sector (for example, for Sector 23: \url{https://tasoc.dk/docs/release_notes/tess_sector_23_drn32_v03.pdf}).} close in time of the transits of \plaa, \plab\, and \plac, and we therefore rule these types of instrumental systematic scenarios out.

\subsection{\texttt{DAVE} validation}

We conducted a uniform vetting analysis of the three candidates in addition to the SPOC data validation reports using the pipeline called discovery and vetting of exoplanets (\texttt{DAVE}; \citealt{Kostov2019}), The pipeline consists of different modules enabling a thorough analysis of transit events on two levels: an assessment at the pixel level through photocenter analysis (\texttt{centroid}), and an evaluation of the LC at the flux time-series level (\texttt{Modelshift}).

The \texttt{centroid} module constructs a differential image by subtracting the in-transit image from the out-of-transit image. It determines the centroid by fitting the \tess\ pixel response function with the image, and this process is carried out for each detected transit. Subsequently, it computes the mean centroid position across occurrences, along with its statistical significance, which contributes to the vetting process. The aim of this module is to determine whether the source of the transit is the target star or some nearby stars. Any deviation in the  position of the centroid from the location of the target star is an indication of a false-positive (FP) event. The module \texttt{Modelshift} generates a phase-folded LC using the optimal trapezoidal transit model. Its principal aim is to determine whether the signal source corresponds to an eclipsing binary system. This module provides some transit parameters, such as the mean primary signal and other prominent features. Furthermore, it contrasts odd and even transits to gauge their statistical differentiation, and it evaluates the transit shape, which provides a thorough evaluation of the observed transit characteristics.

We vetted \plaa\ transit events in \tess\ Sectors 14,20,40,47,53 and 60, \plab\ in Sectors 42, 43, and 70, and \plac\ in Sectors 21 and 48. \plaa\ and \plab\ passed all the tests as a bona fide planet candidate. 
We report as an example the \texttt{centroids} module output for the two planets in Fig.~\ref{fig:1243_centroids}. The photocenter of both planets is close to the position of the target, despite a negligible offset due to contamination of nearby resolved sources. 
The brightest pixel in the difference image perfectly corresponds to the target positions.  The Lomb-Scargle periodogram revealed no modulation of the LCs of the two stars that might be attributed to ellipsoidal variations typical of a close binary star system.
We also report the output of the module \texttt{Modelshift} for \plaa\ and \plab. \texttt{Modelshift} returns a clear primary transit for both planets, well above the noise level without any statistically significant difference between odd and even transits. Moreover, there is no evidence of a secondary feature that might have been caused by the occultation of an eclipsing binary. 

In contrast to the previous cases, \plac\ is a low S/N event. The results of the module \texttt{centroid} were unreliable because it is triggered by the strong variability of a bright background binary star at about 1.7\arcmin\ to the northeast (BD+36~2033, \tess\ magnitude $T \approx$ 9.73\,mag), which is outside the aperture mask. Nonetheless, we were still able to discern a brightness variability during the time of transit in correspondence of the target position. The Lomb-Scargle periodogram found no clear evidence of sinusoidal variations in the LC. Despite some flags raised by the low S/N of the event, we classified the transit as a bona fide planet candidate. To ensure thoroughness, we opted to perform a pixel-level LC (PLL) analysis alongside the \texttt{centroid} module. This analysis enabled us to examine the LC for every pixel within the field of view of the associated \tess\ pixel files, which verified whether the transit occurred in adjacent pixels of the mask. 
This additional layer of scrutiny has proven to be very useful when \texttt{DAVE} photocenter measurements were unreliable or difficult to interpret \citep{Magliano2023}. Fig.~\ref{fig:PLL_TOI5388} shows the PLL analysis for \plac\ in a 3-day window during the fifth transit in \tess\ Sector 21. 
This plot shows that the LC associated with the \tess\ pixel file shows a shallow transit event. This variability is obscured by the modulation of the brighter source BD+36~2033 at the time of the transit, however. To ensure that the transit signal variability was not caused by this nearby star, we downloaded the \tess\ data for Sectors 21 and 48 for this target. The GLS periodograms revealed no periodicity that could be correlated with the 2.59\,day transit signal.
For completeness, we also repeated the PLL analysis for \plaa\ and \plab and thus confirmed the results obtained by the module \texttt{centroid}.

\subsection{Flux contamination by nearby sources}

The large pixel scale of\tess of approximately $21''$\,pixel$^{-1}$ and the focus-limited point spread function can lead to flux contamination in individual pixels from nearby or  background and foreground sources. Even though the \texttt{centroid} module shows no offset in the \tess\ image, additional sources within the $21''$ wide pixel may contaminate the transit event or produce a transit-like signal. In the former case, contamination from unresolved sources might bias the transit depth, resulting in underestimated radii; in the latter (and worse) case, the target star is not the source of the transit. To investigate possible contamination from resolved and unresolved stars, we consulted the Gaia stellar catalog \citep{GaiaColl2023} to determine whether any sources were within the aperture mask.
    
\splaa\ ($T$ = $11.20$\,mag) has three dimmer resolved blends within the \tess\ aperture mask: TIC 219698777 ($T$ = $16.52$\,mag), TIC 219698778  (2MASS J09025843+7137420, $T$ = $13.83$\,mag), and TIC 802624803 ($T$ = $18.64$\,mag), which are at a distance of 25\arcsec, 32\arcsec, and 49\arcsec\ . The brightest background companion is 2.5\,mag fainter than the target, which slightly contaminates the flux and alters the result from the module \texttt{centroids}, as shown in Fig.~\ref{fig:1243_centroids}.

\splab\ ($T$ = $10.14$\,mag) has five dimmer resolved blends, but the brightest source, TIC 10002460038, is almost 7\,mag fainter than the object. In this case, the flux contamination due to nearby targets is completely negligible. 

Last, \splac\ ($T$ = $10.57$\,mag) has only two fainter resolved blends, and the brightest, TIC 407591299 ($T$ = $15.61$\,mag), is five magnitudes dimmer than the object. As in the previous cases, we neglected the photometric contamination of the nearby stars.

\subsection{On-source transit confirmation and statistical validation}

We also independently validated the three candidates statistically using the \texttt{TRICERATOPS} pipeline \citep{Giacalone2021} to evaluate the probability of FP scenarios. We ran \texttt{TRICERATOPS} using the phase-folded LCs on the orbital periods obtained by the photometric modeling. We constrained the overall calculation with the results of the high-contrast imaging. 
Using ground-based high-contrast imaging observations, we detected the source of a transit signal on-target, with nearby FP probability (the probability that the signal arises from a nearby resolved star) to be equal to zero. Thus, we excluded other nearby sources, which clearly do not contaminate the transit signal, from our false-positive probability (FPP) calculations.
We obtained FPP $=0.010 \pm 0.001$ and $=0.017 \pm 0.001$ for \plab\ and \plac, which statistically validates them as transiting exoplanets. 
For \plaa, the obtained FPP is highly sector dependent and not very reliable. Moreover, some nearby targets contaminate its primary flux in this analysis, as shown above, which makes the \texttt{TRICERATOPS} validation challenging. We obtained a final FPP $\sim 0.20$, and based on this analysis, we were therefore not in a position to consider it statistically validated. Based on the results of the previous photometric analyses and considering the clear RV detection of the planet (Sect.~\ref{section:data_analysis}), however, we confirmed the planetary nature of this \plaa.

\section{The stars}
\label{sec:star}

\subsection{Stellar parameters}
\label{section:stellar_parameters}

\begin{table*}
\centering
\caption{Sellar parameters of the three planet-host stars}
\begin{tabular*}{1.7\columnwidth}{lcccl}

\hline
\hline

\multicolumn{1}{l}{Parameter} & \multicolumn{1}{c}{\splaa} & \multicolumn{1}{c}{\splab} & \multicolumn{1}{c}{\splac}  & \multicolumn{1}{l}{Reference$^{(\mathrm{a})}$}\\

\hline

\multicolumn{5}{c}{Main identifiers and spectral type} \\[0.5ex]
Discovery name & LSPM J0902+7138 & G 2--21 & Wolf 346 & Lep05, Gic59, Wol19 \\
Karmn & J09029+716 & J01078+128 & J10087+355 & AF15 \\
TIC    & 219698776 & 384888319 & 407591297  & Sta18/19 \\
Spectral type     & M2.0: V & M1.5 V & M3.0 V & This work, Lep13, Rei04 \\[1ex]

\multicolumn{5}{c}{Astrometry} \\[0.5ex]
$\alpha$ (J2000) & 09:02:55.82 &  01:07:52.53 & 10:08:42.37 & {\em Gaia} DR3 \\
$\delta$ (J2000) & 71:38:11.3 & 12:52:51.4 & 35:32:51.2 & {\em Gaia} DR3 \\
$\mu_\alpha \cos{\delta}$ ($\mathrm{mas\,a^{-1}}$) & $-100.482 \pm 0.013$ & $275.861 \pm 0.036$ & $192.380	 \pm 0.026$ & {\em Gaia} DR3 \\
$\mu_{\delta}$ ($\mathrm{mas\,a^{-1}}$) & $-142.980 \pm 0.016$ & $-22.973 \pm 0.025$ & $133.239 \pm 0.021$ &  {\em Gaia} DR3 \\
$d$ (pc) & $43.126 \pm 0.029$ & $28.384 \pm 0.023$ & $18.5538 \pm 0.0092$ & {\em Gaia} DR3 \\[1ex]

\multicolumn{5}{c}{Photometry$^{(\mathrm{b})}$} \\[0.5ex]
$B$ (mag)    & $14.923 \pm 0.100$         & $13.601 \pm 0.033$       & $14.390 \pm 0.011$        & APASS9, UCAC4, APASS9 \\
$V$ (mag)    & $13.427 \pm 0.037$       & $12.214 \pm 0.146$       & $12.728 \pm 0.064$       & APASS9, UCAC4, APASS9 \\
$G$ (mag) & $12.3452 \pm 0.0028$    & $11.2167 \pm 0.0028$    & $11.7206 \pm 0.0028$   & {\em Gaia} DR3 \\
$T$ (mag) & $11.1966 \pm 0.0074$ & $10.1406 \pm 0.0073$ & $10.5737 \pm 0.0073$ & Sta18/19 \\
$J$ (mag)    & $9.731 \pm 0.018$        & $8.787 \pm 0.020$         & $9.167 \pm 0.018$        & 2MASS \\
$H$ (mag)    & $9.093 \pm 0.017$        & $8.175 \pm 0.017$        & $8.618 \pm 0.029$        & 2MASS \\
$K_s$ (mag)    & $8.878 \pm 0.016$        & $7.954 \pm 0.024$        & $8.319 \pm 0.021$        & 2MASS \\[1ex]

\multicolumn{5}{c}{Photospheric parameters} \\[0.5ex]
$T_{\rm eff}$  (K)   & $3515 \pm 79$ &  $3697 \pm 71$  &  $3440 \pm 85$  & This work \\
$\log{g}$ (cgs)  & $4.73 \pm 0.12$ & $4.61 \pm  0.07$ &  $4.99 \pm  0.21$& This work\\
{[Fe/H]} (dex)   & $-0.20 \pm 0.16$ & $-0.24 \pm 0.07$ & $-0.36 \pm 0.20$  & This work\\[1ex]

\multicolumn{5}{c}{Physical parameters} \\[0.5ex]
$L_{\rm bol}$ ($10^{-4}~\mathrm{L_\odot}$) & $358.5 \pm 7.2$ &  $386.7 \pm 7.0$ & $113.40 \pm 0.51$ & This work \\
$R_\star$ ($\mathrm{R_{\odot}}$) & $0.511 \pm 0.024$ & $0.480 \pm 0.019 $ & $0.300 \pm 0.015$ & This work \\
$M_\star$ ($\mathrm{M_{\odot}}$) & $0.515 \pm 0.027$ & $0.482 \pm 0.023$ & $0.292 \pm 0.018$ & This work \\ 
$\rho_\star$ ($\mathrm{g \, cm^{-3}}$)   &  $5.4 \pm 0.8$ & $6.1 \pm 0.8$ & $17 \pm 3$ & This work\\[0.5ex]

\multicolumn{5}{c}{Stellar activity} \\[0.5ex]
$v\,\sin i$  ($\mathrm{km\,s^{-1}}$)  & $<2$ &  $<2$  & $<2$ & This work \\
$P_{\rm rot}$ (d)   & ... & $20.7 \pm 1.4$ $^{(\mathrm{c})}$ & ...  & This work\\

\hline
\end{tabular*}
\tablefoot{$^{(\mathrm{a})}$ The references cited on this table are 
Lep05: \citealt{stellar_catalogue_2005}, 
Gic59: \citealt{1959LowOB...4..136G},
Wol19: \citealt{1919VeHei...7..195W},
AF15: \citealt{catalogue_M_dwarfs}, 
Sta18/19: \citealt{TessInputCatalog2018,stassun2019}, 
Lep13: \citealt{lepine2013}, 
Rei04: \citealt{reid2004}, 
{\em Gaia} DR3: \citealt{GaiaColl2023}, 
APASS9: \citealt{carmeness_2014},
UCAC4: \citealt{ucac4},
2MASS: \citealt{2mass_2016}. 
$^{(\mathrm{b})}$ We also compiled photometry in the GALEX $NUV$, SDSS9 $u'\,g'\,r'\,i'$, UCAC4/APASS $r'\,i'$, and WISE $W1\,W2\,W3\,W4$ bands \citep{cif20}.
$^{(\mathrm{c})}$ Rotational period derived from the weighted average of periodicities in the spectroscopic activity indicators, photometric data, and the GP hyperparameter $P_{\rm rot}$.}
\label{tab:stellar_parameters}
\end{table*}

The stellar atmospheric parameters of the targets, namely effective temperature $T_{\rm eff}$, surface gravity $\log{g}$, and iron abundance [Fe/H], were derived from the high S/N VIS and NIR CARMENES template spectra generated by {\tt SERVAL} with the publicly available package {\tt SteParSyn}\footnote{\url{https://github.com/hmtabernero/SteParSyn/}} \citep{tab22}. The {\tt SteParSyn} parameter determination used a synthetic model grid computed with BT-Settl \citep{all12} models as described by \cite{marf21}. 
In addition, the rotation velocity $\varv \sin{i}$ of our targets, which is a {\tt SteParSyn} input parameter, were determined using the method of \citet{rei18} and are $\leq$~2\,\kms\ in all instances. The bolometric luminosity was computed from the integration of the spectral energy distributions from Johnson $B$ to WISE $W4$ as \citet{cif20} with updated \textit{Gaia} DR3 parallaxes and photometry. These parameters were later used with the procedures described by \citet{sch19} to infer the stellar mass and radius.  The spectral type of \splaa\ (LSPM J0902+7138) was estimated from photometry with the relations of \citet{cif20}, while spectral types of \splab\ (G~2--21), \splac\ (Wolf 346) were measured from low-resolution optical spectroscopy. The three stars are all early-M dwarfs (M2.0\,V, M1.5\,V, and M3.0\,V). The full list of stellar parameters from the literature with the corresponding references and the parameters derived in this work are reported in Table~\ref{tab:stellar_parameters}. The three stars also have no common proper motion and parallax companions down to the \textit{Gaia} DR3 magnitude completeness at separations greater than 0.4\arcsec\ \citep{cifuentes2025}, and they belong to the Galactic thin-disk kinematic population, but are not members in any known stellar kinematic group \citep{cortescontreras2024}.

\subsection{Activity indicators and stellar photometry analysis}
\label{ssec:activity_indicators}

It is crucial to study the activity indicators for detecting periodic behavior in the RV time series and distinguishing between planetary signals and stellar activity. The algorithm called generalized Lomb-Scargle periodogram (GLS; \citealt{zechmeister2009}) is used to detect and characterize periodic signals in unevenly sampled data. 
For \splaa, \splab, and \splac, we performed a GLS periodogram analysis on the RV data, stellar activity indicators, \tess\ SAP photometry, and long-term photometric data. The periodograms were computed using the package \texttt{PyAstronomy}\footnote{\url{https://github.com/sczesla/PyAstronomy}}  \citep{pyastronomy2019}. In the long-term photometric data, we initially performed a linear fit to the data to exclude any long-term variations (i.e., due to proper motions) that would prevent the detection of the rotation period signal.
We excluded the \tess\ SAP data around predicted transit midtimes ($\pm 1$ transit duration from the midtime) to avoid affecting the possible detected periodicity.  
Moreover, because periodograms are sensitive to outliers, we rejected outliers in the \tess\ and ground-based data using an uncertainty-weighted average rolling window of 10 points. Data points whose distance from the window mean was larger than the square sum of the rolling window standard deviation and the data-point uncertainty were removed.

To assess the significance of the detected signals, we calculated the false-alarm probability (FAP). A signal was considered statistically significant with an $\mathrm{FAP < 0.13\%}$. To compute the FAP, we used a bootstrap approach following \cite{kokori2022, kokori2023}. The full periodograms of the spectroscopic activity indicators and long-term photometric data can be found in Appendix~\ref{app:periodograms}.

For \splaa, the GLS periodograms of the spectroscopic activity indicators (Fig.~\ref{fig:per1243}) reveal no significant peaks in the range of 2--200\,d. Neither the \tess\ SAP LCs nor the ground-based LCs (LCOGT-$B$, LCOGT-$V$, and e-EYE-$B$) of \splaa\ exhibit a dominant photometric modulation. Most of the periodogram peaks of the \tess\ data are very close to the observational baseline covered by the data. Most of the peaks of the ground-based data are on the order of a few days, which shows that they are most likely affected by the data sampling. No robust conclusions can be drawn regarding the rotation period of \splaa.

For \splab, the analysis identified a significant activity signal at $\sim 20.5$ days (FAP $< 0.13\%$) in the H$\alpha$ and Ca~{\sc ii}~IRT$_1$, Ca~{\sc ii}~IRT$_2$, and Ca~{\sc ii}~IRT$_3$ activity indicators. This signal also appears in the RV data, but with a slightly longer period of 21.4\,d (Fig.~\ref{fig:per4529}). There is a large gap of about 200\,d in the observations, however, which can affect the result of the periodogram analysis. When the periodogram was run on only the second part of the data, which is the largest sample, the periodogram peak of the RV data was at 20.4\,d, which coincides with the maximum periodicities found in the activity indicators. When the periodograms were run only for the second part of the data, the periodogram peaks did not shift by more than 0.1\,d for any activity indicator. The final derived rotational period from the spectroscopic activity indicators (taken from the periodograms of the full dataset) was found by the weighted average of individual activity indicators and has a value of $20.5 \pm 1.0\,\mathrm{d}$, where we calculated the uncertainty by fitting a Gaussian function on the periodogram peak in the frequency space. 

The long-term ground-based LCOGT photometry of \splab\ (Fig.~\ref{fig:4529longterm}) shows a statistically significant peak at $20.5\pm 2.0 \,\mathrm{d} $ in the $B$ band and two statistically significant peaks at $10.0\pm 1.0 \,\mathrm{d}$ and $19.5 \pm 1.5 \,\mathrm{d}$ in the $V$ band. The $10 \,\mathrm{d}$ peak is most likely an alias of the rotation period because it is not evident in any other photometric or spectroscopic indicator. TJO shows significant peaks at $12 \pm 1.5\,\mathrm{d}$ and $58 \,\mathrm{d}$, whereas e-EYE data display no significant periodicities.

The \tess\ SAP flux from Sectors 42 and 43 of \splab\ shows significant peaks at $17.0 \pm 2.0\,\mathrm{d}$ when analyzed individually, and Sector 70 shows a peak at $18 \pm 2\,\mathrm{d}$. To improve our analysis, we combined the two consecutive sectors by normalizing the flux and adding an offset to data from Sector 43 to take the flux sector jumps into account. The offset was found by fitting it simultaneously with a sinusoidal curve to both sectors, and it has a value of $0.0043$ in our normalized flux units (Fig.~\ref{fig:4529longterm}). The combined \tess\ sectors again show a significant peak at $18.5 \pm 2.0 \,\mathrm{d}$, which is more reliable than the individual sector analysis because it covers a longer time span. The final adopted rotational period from the photometric measurements comes from the weighted average of \tess\ data, ground-based photometry, and has a value of $19.5 \pm 2.0 \,\mathrm{d}$.
To account for the stellar activity and to improve the planetary signal retrieval, we modeled the stellar activity using a quasiperiodic Gaussian process (GP) kernel in our RV fit (Sect.~\ref{ssec:joint_analysis}). In the RV fit, we adopted a prior on the stellar rotation period of $20 \pm 3 \,\mathrm{d}$, which enabled us to determine a GP rotational period hyperparameter of $20.9 \pm 0.8 \,\mathrm{d}$ for \splab\ (Sect.~\ref{ssec:joint_analysis}). Finally, using the combination of the spectroscopic activity indicators, the photometric measurements, and the Gausian process fitting, we derived our final adopted rotation period at $20.7 \pm 1.4 \,\mathrm{d}$.

To conclude, using the rotation period of \splab\,, we were able to estimate the age of the star using gyrochronology. The results may differ depending on the exact relation used (\citealt{barnes2007, mamajek2008, agnus2015}). We adopted a final gyrochronology age of $1.0^{+1.0}_{-0.5}$\,Gyr.

The GLS periodogram of the spectroscopic data reveals no significant peaks for \splac  (Fig.~\ref{fig:per5388}). The \tess\ SAP LC shows some evidence of significant peaks at about 15\,d. The \tess\ SAP flux shows flux jumps, however, and its observing window is about 30\,d with a gap in the middle. Furthermore, the \tess\ orbital period for Sector 48 was close to 15\,d, which might explain the photometric variations at this period. The background light often has a strong contribution from the reflected light from the Earth, which exhibits diurnal-like variations. Taking all the above and the absence of a 15\,d periodicity in any other photometric or spectroscopic indicator into account, we concluded that the 15\,d periodicity is most likely an instrumental feature of \tess\ and not evidence of periodic photometric variation of \splac.
The 40d peak of the TJO data is intriguing, especially because during the time period when TJO observations overlap the LCOGT observations, we saw a similar flux gradient. More data are required to draw robust conclusions, however, and since this periodicity is completely absent in the RV data, we omitted stellar activity modeling in the RV analysis.

In summary, for \splaa\ and \splac\, no clear periodic activity modulation was identified. For \splab\, the analysis supported an evident stellar activity with a rotational period of $20.7 \pm 1.4 \,\mathrm{d}$.

\section{Data analysis}
\label{section:data_analysis}

\subsection{Transit LC analysis}
\label{ssec:lightcurve_analysis}

We modeled the transits LCs using the package \texttt{Pylightcurve}\footnote{\url{https://github.com/ucl-exoplanets/pylightcurve}} \citep{pylightcurve}. The limb-darkening coefficients for each planet were calculated using the package \texttt{ExoTETHyS}\footnote{\url{https://github.com/ucl-exoplanets/ExoTETHyS}} \citep{Morello2020a, Morello2020b}. For all the instruments, we calculated quadratic limb-darkening coefficients using their equivalent bandpasses, and we used them in the LC fitting. We employed the initial ExoFOP planetary parameters to select the best LCs to include in the global photometric fit and in the joint photometric and RV fit. To do this, we computed the S/N for each transit. After evaluating these S/N values and visually inspecting the LCs, we adopted all the \tess\ LCs for the three planets, while for the ground-based LCs, we only used the full transits with a S/N higher than the median \tess\ S/N for each planet. For \plaa, all the ground-based LCs passed our selection criteria, for \plac\, we selected the LCOGT and SAINT-EX LCs, and for \plab\, we omitted all the ground-based transits.

Before we proceeded with the LC fit, we removed systematic trends from all the selected LCs using a second-order polynomial with time. Our choice to detrend the LCs with polynomials instead of employing more complex modeling (e.g., GP) for modeling the LCs allowed us to keep uniformity in our analysis, minimize the introduction of unnecessary biases, and reduce the computation time. For each planet, we fit the period $P$, transit midtime $T_0$, planet-to-star radius ratio $R_{\rm p}/R_{\rm s}$,
inclination $i$, and scaled semimajor axis $a$. For all planets, we assumed a null eccentricity and $90 \degree$ argument of periastron.
We performed the fit using the MCMC package \texttt{emcee}\footnote{\url{https://github.com/dfm/emcee}} \citep{emcee}. The number of chosen walkers was three times the number of fitted parameters, and we ran the chains until convergence, applying the convergence criteria described by \cite{fulton2018}.

Since fitting LCs collectively with a fixed period and a unique $T_0$ is not suited for detecting transit time variations (TTVs) and may result in smoothing out variations in the planetary parameters, we also performed a TTV analysis.
In this case, we used all available LCs, fit individually, in order to first verify the robustness of our findings, and second, to fine-tune the planetary parameters. For the TTV check, we optimized the parameters $T_0$ and $R_{\rm p}/R_{\rm s}$, and we also applied a quadratic detrending polynomial, assuming broad uniform priors for all fit parameters. This method allowed us to search for any significant deviations from a linear ephemeris and for variations in $R_{\rm p}/R_{\rm s}$. The fit to each individual ground-based LC can be found in Appendix \ref{appendix:ground_based_lcs} (Figs.~\ref{fig:g1243_1}, \ref{fig:g4529}, and \ref{fig:g5388}).

The outcome of this analysis was the observed (O) minus calculated (C) plots for the three planets, which were produced in a similar way to \cite{kokori2022, kokori2023}. We computed the periodogram of these O-C plots and fit a line to determine whether any periodic TTV components or long-term linear trends could be identified. None of the planets shows evidence of TTVs. The analysis for \plab\ indicates, however, that the SAINT-EX and the MuSCAT2 transit taken on 2022\,October\,30 are outliers because the S/N of these transits is low, the error bars are underestimated ,and some are partial. These reasons make it difficult to determine a precise transit midtime from the single-transit fit.

\subsection{RV analysis and joint fit}
\label{ssec:joint_analysis}

We computed the GLS periodograms of the RV data to search for periodic patterns and determine whether we might identify the planetary signals in the RV time series. 
For \splaa, the maximum peak of the RV periodogram (Fig.~\ref{fig:per1243}) is at $10.2$~d, which is about twice the expected $4.66$ d transit signal period, and it is also present in the periodogram. Both peaks have a relatively high FAP of 27\% and 39\%, respectively. The $10.2$ d peak is not detected in any of the activity indicators and it is therefore unlikely to be of stellar origin. We therefore investigated the possibility of a planetary origin. The orbital models for the \splaa\ system including two planets were not favored with respect to the one-planet model, however, suggesting that it is most likely an alias of the actual planetary period. Finally, there is no additional significant signal in any other activity indicators.

The analysis of \splab, shown in Fig.~\ref{fig:per4529}, highlights stellar activity, as discussed in Sect.~\ref{ssec:activity_indicators}. 
In the RV data, the most statistically significant peak is identified at around 20--21\,d, corresponding to the rotational period of the star. The transit period at $5.88$\,d is not clearly seen in the periodogram, as expected given the prevalence of the stellar activity, which needs to be carefully treated to disentangle the planetary signal. 

As Fig.~\ref{fig:per5388} shows, the $2.58$ d transit period of \plac\ is absent in the RV periodogram. This is expected because the planetary radius is small ($\sim 1$ R$_{\oplus}$) and the RV semi-amplitude of the planet signal is therefore correspondingly small ($\sim 0.5\, \mathrm{m\,s^{-1}} $) as predicted using the known planet radius, orbital period, and stellar mass. For \splac, the median internal precision of our dataset is $2.2\,  \mathrm{m \, s^{-1}}$, and the RMS is $4.2\,  \mathrm{m \, s^{-1}}$, meaning that with our 42 RVs we would be able to detect a $\mathrm{3 \sigma}$ amplitude of $1 \,  \mathrm{m \, s^{-1}}$ only in the best case.
This expected RV amplitude in combination with the instrument uncertainty prevents a precise mass determination from the RV fit with the current dataset, but it allowed us to determine an upper limit for the mass. 

To model the Keplerian signals, we used the \texttt{RadVel}\footnote{\url{https://github.com/California-Planet-Search/radvel}} package \citep{fulton2018}, as implemented in the \texttt{juliet}\footnote{\url{https://github.com/nespinoza/juliet}} \citep{juliet_2019} framework. 
For \plaa\ and \plac, we fit the semi-amplitude $K$, a detrending polynomial with reference epoch at the mean time of the observing window, which is characterized by $\gamma$ (a constant offset), $\dot{\gamma}$ (a linear trend), $\ddot{\gamma}$ (a quadratic trend, with reference epoch at ${T_{\rm ref} = 2\,458\,460}$), and a jitter term $J$ (the jitter term accounts for additional noise such as stellar activity or instrumental noise, which is not captured by the measurement uncertainties). The period and transit midtimes were treated as free parameters with tightly constrained Gaussian priors derived from the posterior distributions in our transit analysis. The eccentricity and argument of periastron were kept constant, assuming a circular orbit. We carried out a nested sampling analysis with package \texttt{dynesty}\footnote{\url{https://github.com/joshspeagle/dynesty}} \citep{dynesty_2019}, employing 500 live points and adopting $\Delta \log \mathcal{Z} < 0.1$ as the convergence criterion, where $ \log \mathcal{Z}$ is the natural logarithmic Bayesian evidence.

Since \splab\ shows strong evidence of stellar activity, we modeled it including GPs in the fit. We used the quasiperiodic (QP) kernel as implemented in \texttt{celerite}\footnote{\url{https://github.com/dfm/celerite}}  \citep{celerite2017}, which has been shown to be effective in modeling stellar activity (e.g., \citealt{Rajpaul2015, nickolson_2022, stock_2023}), and has been widely used in the literature for mass determination studies. This kernel was first introduced by \cite{celerite2017} as a more computationally efficient kernel with respect to the classical QP kernel (i.e., \citealt{Rajpaul2015}). The hyperparameters of the kernel are the GP amplitude (${B_{\rm GP}}$), which controls the scale of the GP variations, the GP amplitude unitless multiplicative factor (${C_{\rm GP}}$), which affects the GP amplitude, the exponential length scale of the GP (${L_{\rm GP}}$), which controls how quickly the correlation between the data decreases over time, and the rotational period of the GP (${P_{\rm rot}}$), which describes the GP periodic component.
Table \ref{tab:model_comparison} shows that the QP GP model is strongly favored with respect to the other models that we tested (flat line, Keplerian, Keplerian + trend, and Keplerian + sinusoidal model for activity), with $\Delta \log \mathcal{Z} > 2$ \citep{kass&raftey1995}. The main effect on our final results is that if we had not used a GP, the upper limit on the semi-amplitude would increase, resulting in a more conservative estimate of the upper mass limit. 
For \plab, we therefore fit the period and transit midtime (assuming Gaussian priors from the photometric fit), semi-amplitude $K$, and jitter $J$, together with the GP hyperparameters, as explained above. We adopted uniform priors for ${B_{\rm GP}}$, ${C_{\rm GP}}$, and ${L_{\rm GP}}$. For ${P_{\rm rot}}$, we assumed a Gaussian prior with broad variance based on the value derived from the activity indicators and long-term photometry analysis $\mathrm{(\mu = 20.5\, d, \, \sigma ^2 = 3})$.
As in the previous cases, we used a nested sampling analysis to find the posterior distribution of our model, using 500 live points, and adopting $\Delta \log \mathcal{Z} < 0.1$ for the chain convergence criterion. The final model, including the Keplerian orbit and the detrended model, is shown in Fig.~\ref{fig:rv_gp_4529}.

Finally, we performed a joint RV and LC modeling for a complete characterization of the properties of the exoplanets using \texttt{juliet}. We fit the same photometric parameters as described in Sect.~\ref{ssec:lightcurve_analysis}, except for the inclination, which was replaced by the impact parameter $b$ because the \texttt{juliet} package uses the \texttt{batman} \citep{batman2015} implementation of the transit model, and we assumed the RV model previously described in this section for each planet. 

\begin{figure}[h]
    \centering
    \includegraphics[width=0.45\textwidth]{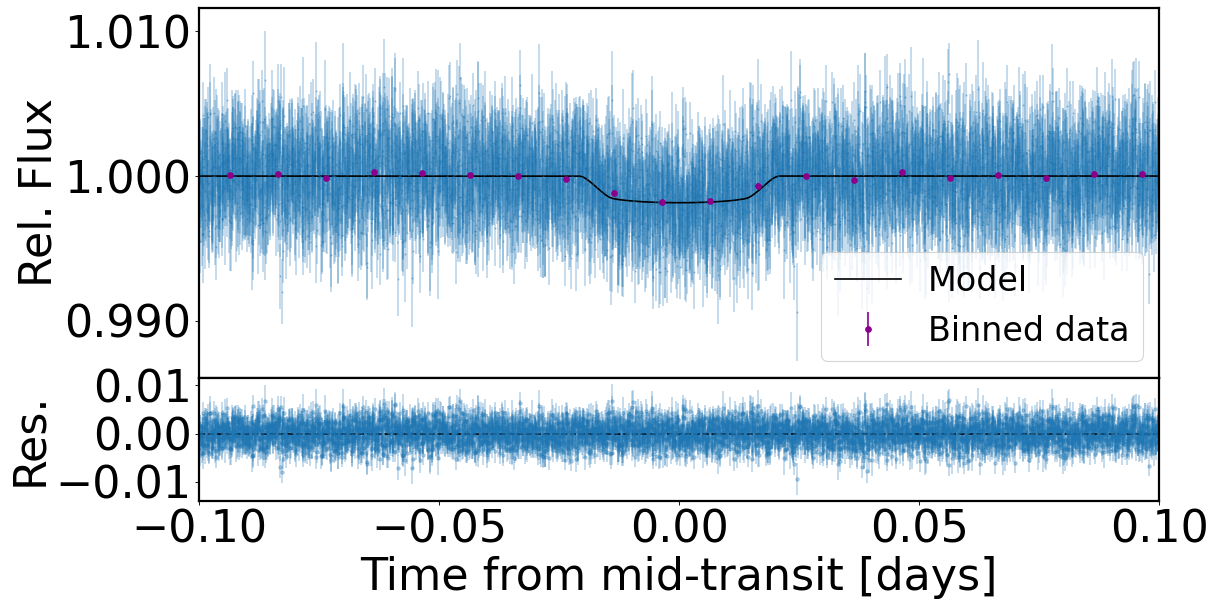}
    \par
    \centering
    \includegraphics[width=0.45\textwidth]{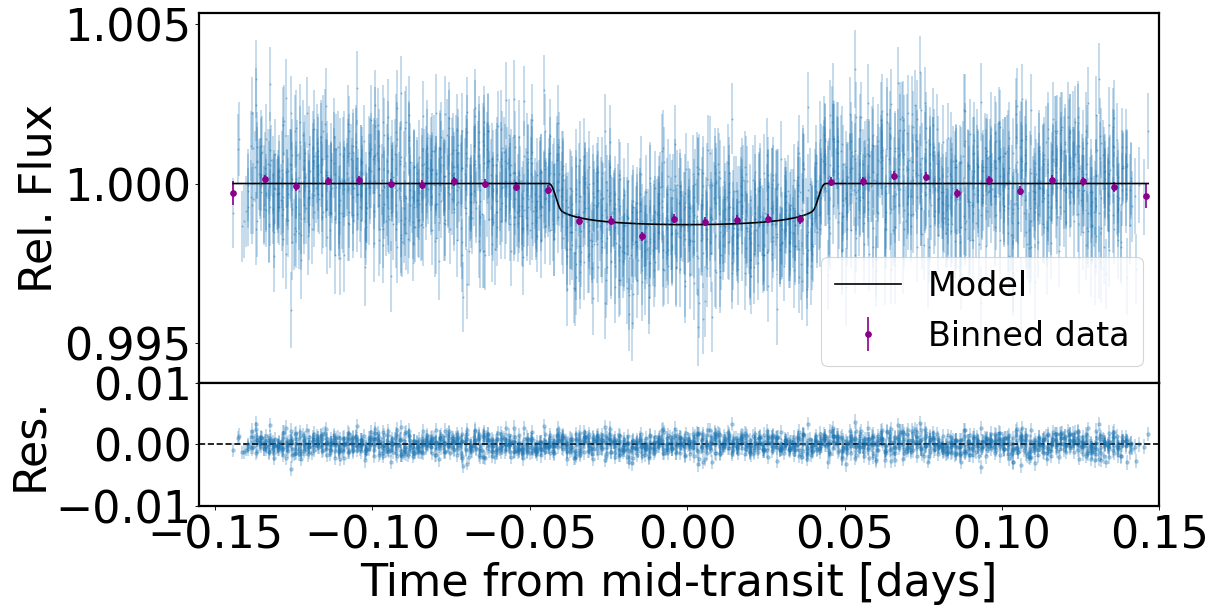}
    \par
    \centering
    \includegraphics[width=0.45\textwidth]{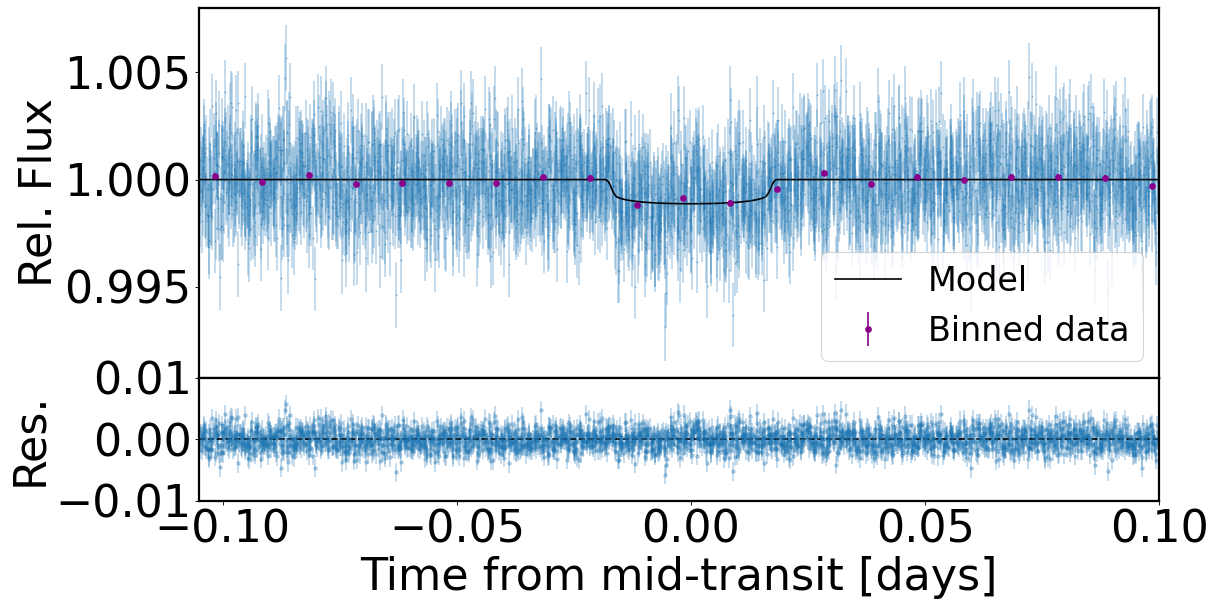}
    \caption{\tess\ phase-folded and detrended relative flux of \plaa\ (top), \plab\ (middle), and \plac\ (bottom). The black line shows the best-fit model, and the purple points show the binned flux.}
    \label{fig:tess_binned_all}
\end{figure}

\begin{figure}[h]
    \centering
       \includegraphics[width=0.45\textwidth]{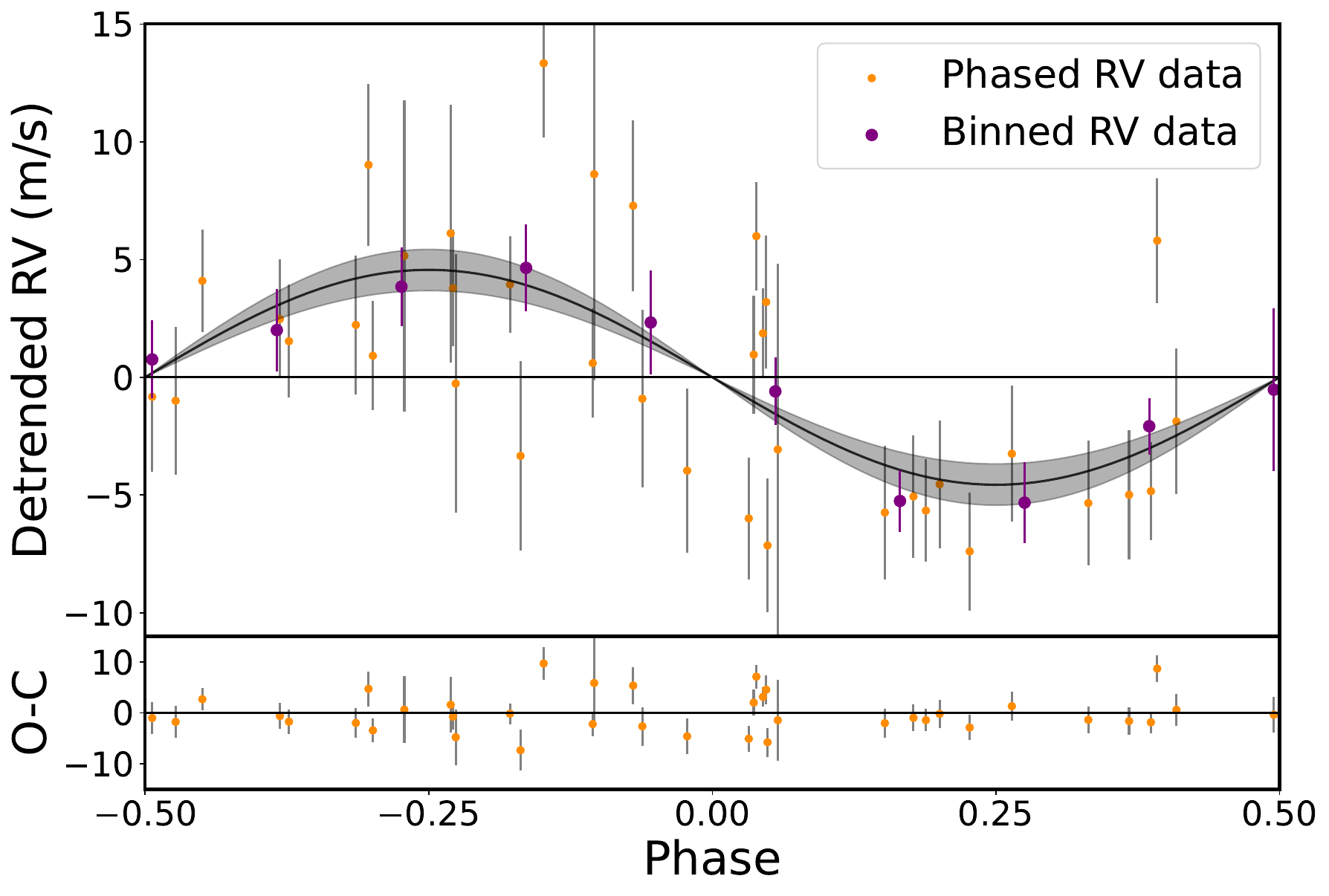}
    \par
    \centering
        \includegraphics[width=0.45\textwidth]{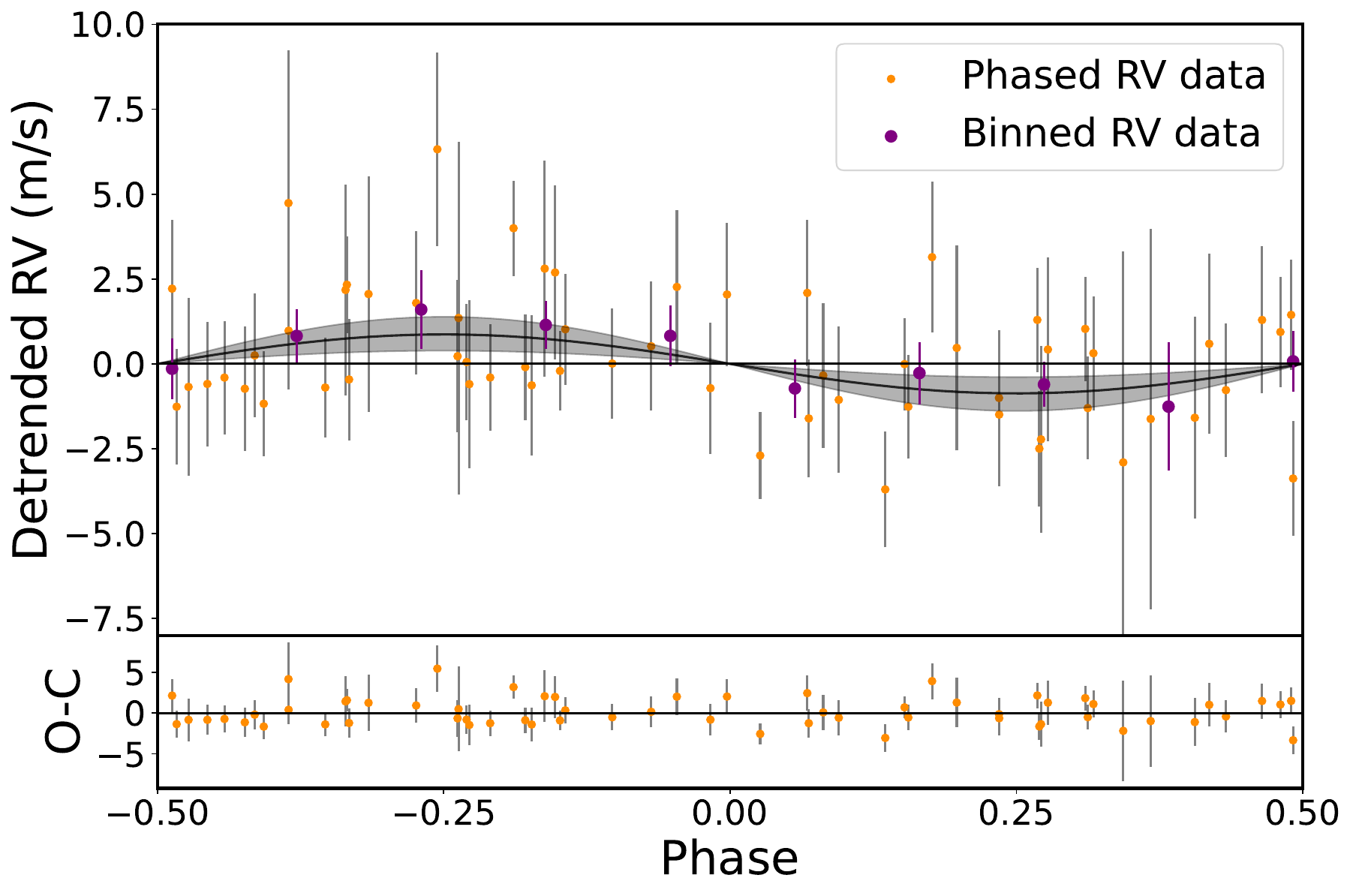}
    \par
    \centering
        \includegraphics[width=0.45\textwidth]{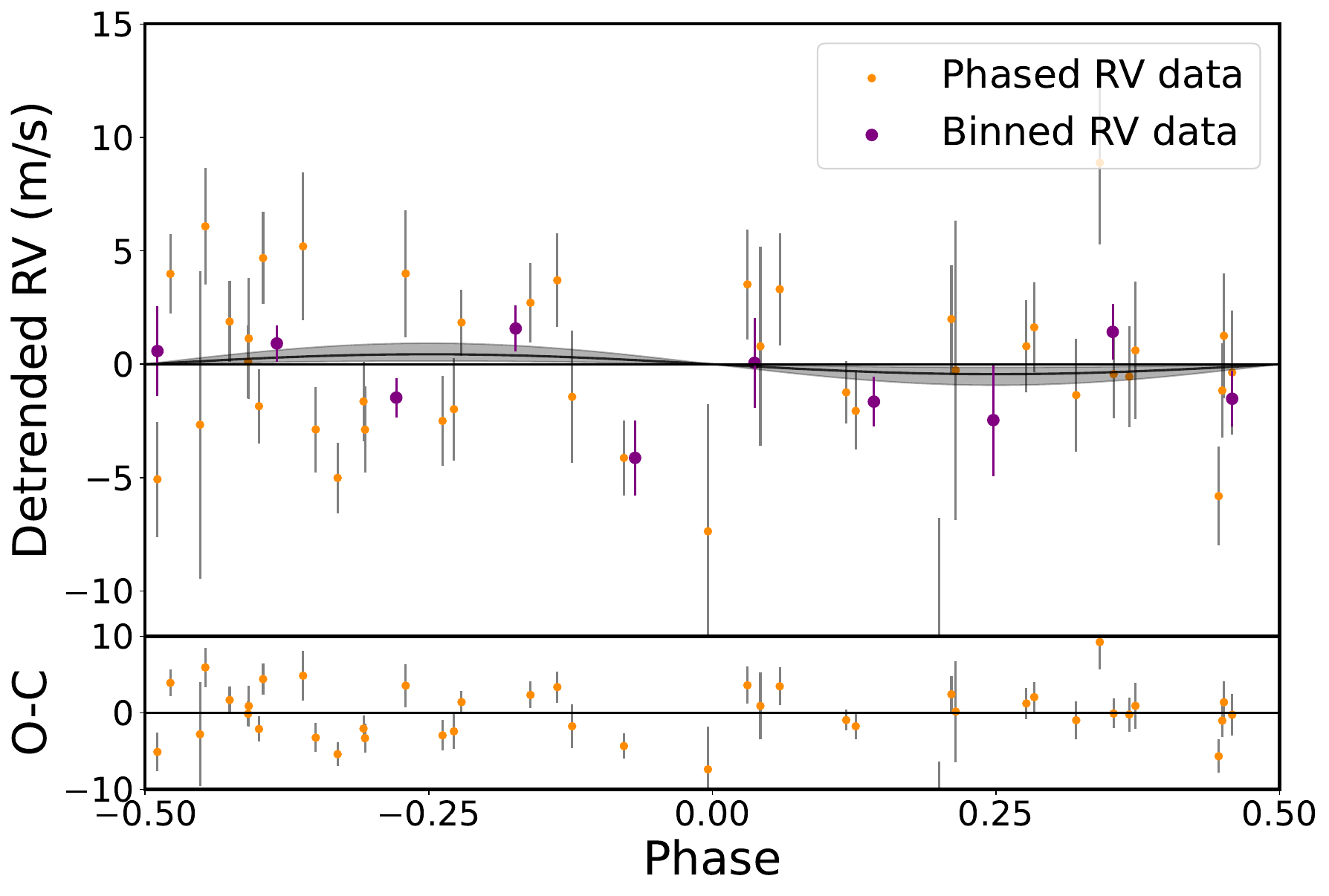}
    \caption{RV phase-folded and detrended plot for \plaa\ (top), \plab\ (middle), and \plac\ (bottom). The purple points show the binned RVs, the black line shows the best-fit model, and gray shaded area shows the 1 $\sigma$ uncertainty of the RV model.} 
    \label{fig:rv_folded}
\end{figure}

We carried out a nested sampling analysis using 1000 live points and applied the same convergence criterion as in the RV analysis. We report the results of the joint fit in Table \ref{table:RESULTS}, and we show the \tess\ LCs and phase-folded RVs in Fig.~\ref{fig:tess_binned_all} and Fig.~\ref{fig:rv_folded}.

\begin{table*}
	\caption{Results of the joint fit models for the planets}
    \centering
		\begin{tabular}{lccc}
        \hline
        \hline
			\multicolumn{1}{r}{Parameter} & \plaa\ & \plab\ & \plac\ \\
            \hline
            & \multicolumn{3}{c}{Fitted parameters}\\[1ex]
            $R_{\rm p} / R_{\rm s}$ & $0.0418^{+0.0011}_{-0.0008}$ & $0.0338^{+0.0009}_{-0.0008}$ & $0.0305^{+0.0017}_{-0.0010}$ \\
			$P$ (d) & $4.6594779^{+0.0000042}_{-0.0000045}$ & $5.879577^{+0.000011}_{-0.000010}$ & $2.5946748^{+0.0000042}_{-0.0000030}$ \\
			$T_0$ (BJD) & $2459369.83103^{+0.00028}_{-0.00027}$ & $2459701.3329^{+0.0006}_{+0.0006}$ & $2459348.07817^{+0.00052}_{-0.00055}$ \\
			${a / R_{\rm s}}\,(-)$ & $25.5^{+1.9}_{-2.3}$ & $21.5^{+1.2}_{-2.9}$ & $20.5^{+1.9}_{-4.6}$ \\
            $b\,(-)$ & $0.78 \pm 0.04 $& $0.32 \pm 0.20$ & $ 0.43 ^ {+0.29}_{-0.28}$\\
			$K$ ($\mathrm{m\,s^{-1}}$) & $4.6\pm0.9$ & $0.95^{+0.58}_{-0.79}$ & $0.44^{+0.48}_{-0.30}$ \\
			$\gamma$ ($\mathrm{m\,s^{-1}}$) & $0.0\pm2.0$ & ... & $-0.4^{+2.2}_{-1.8}$ \\
			$\dot{\gamma}$ ($\mathrm{m\,s^{-1}\, d^{-1}}$) & $0.0000046^{+0.0000027}_{-0.0000028}$ & ... & $-0.0000001^{+0.0000014}_{-0.0000013}$ \\
			$\ddot{\gamma}$ ($\mathrm{m\,s^{-1}\, d^{-2}}$) & $-0.0084\pm0.0053$ & ... & $0.0004^{+0.0030}_{-0.0032}$ \\
			$J$ ($\mathrm{m\,s^{-1}}$) & $2.6^{+0.7}_{-0.6}$ & $1.5^{+0.8}_{-0.9}$ & $2.4^{+0.6}_{-0.5}$ \\
			${B_{\rm GP}}\, (\mathrm{m\,s^{-1}})$ & ... & $19^{+4}_{-5}$ & ... \\
			${C_{\rm GP}}\, (-)$  & ... & $0.3^{+0.8}_{-0.3}$ & ... \\
			${L_{\rm GP}\, (\mathrm{d})}$ & ... & $40^{+34}_{-26}$ & ... \\
            $P_{\rm rot}$ (d) & ... & $20.9 \pm 0.8$ & ... \\
            
        & \multicolumn{3}{c}{Derived parameters}\\[1ex]

			${R_{\rm p}}$ (R$_{\oplus}$) & $2.33\pm0.12$ & $1.77^{+0.09}_{-0.08}$ & $0.99^{+0.07}_{-0.06}$ \\
			$M_{\rm p}$ (M$_{\oplus}$)& $7.7\pm1.5$ &  $< 4.9 \, (3 \mathrm{\sigma})$ &  $< 2.2\, (3 \sigma)$ \\
            $\mathrm{\rho_{p} \, (g \, cm ^{ -3})}$& $ 3.3 \pm 0.9$&  $< 4.8 \, (3 \mathrm{\sigma})$ &  $< 11.9\, (3 \mathrm{\sigma})$ \\
            $\mathrm{\rho_{p}}$ ($\rho_{\oplus}$) & $0.61\pm0.15$ &  $< 0.88 \, (3 \mathrm{\sigma})$ &  $< 2.2 \, (3 \mathrm{\sigma})$\\
            $i$ (deg) & $88.24^{+0.20}_{-0.28}$ & $88.7^{+0.9}_{-0.8}$ & $88.8^{+0.8}_{-1.4}$ \\
			${T_{\rm eq}~({\rm K})}$ $^{(\mathrm{a})}$ & $450^{+22}_{-19}$ & $511^{+34}_{-15}$ & $488^{+54}_{-25}$ \\
			${a/ R_{\rm s}}$ (Keplerian) $^{(\mathrm{b})}$ & $18.42\pm0.32$ & $22.39\pm0.36$ & $17.57\pm0.36$ \\
            \bottomrule

		\end{tabular}
        \tablefoot{$^{(\mathrm{a})}$ Assuming $A_{\rm bond}=0.3$. $^{(\mathrm{b})}$ Semi-major axis of the planet orbits, calculated from the stellar masses using the Kepler equation.}
	\label{table:RESULTS}
\end{table*}

\section{Results and discussion}
\label{sec:discussion}

\subsection{Mass, radius, and interior composition}
\label{sec:mass_det_discussion}

For \plaa\, we found a radius of ${R_{\rm p} = 2.33\pm0.12}$  R$_{\oplus}$ and a mass of $M_{\rm p} = 7.7\pm 1.5$ M$_{\oplus}$ ($5 \sigma$ detection). For \plab\, we found a radius of ${R_{\rm p} = 1.77^{+0.09}_{-0.08} \, {\rm R}_{\oplus}}$ and a $3 \sigma$ mass upper limit of ${M_{\rm p} < 4.9 \,{\rm M}_{\oplus}}$. For \plac\, we found a radius of $R_{\rm p} = 0.99^{+0.07}_{-0.06} \, R_{\oplus}$ and a $3 \sigma$ mass upper limit $\mathrm{M_{\rm p} <2.2 \, M_{\oplus}}$. 

\plac\ was also reported by \cite{hord2024identification} as a statistically validated planet that is inside the emission spectroscopy best-in-class samples. \cite{hord2024identification} reported a planetary radius of $R_{\rm p} = 1.89 \, \mathrm{R_{\oplus}}$, however, which is almost double the value we derived. 
The difference arises because \cite{hord2024identification} adopted the parameters from the ExoFOP page ($R_{\rm p} = 1.89 \, \mathrm{R_\oplus}$). This page mentions that the source of the parameters is the QLP from \tess\ Sector 48, which extracted photometry from the full-frame images at exposure times of $600 \,s$. After the identification of the target in the QLP, the SPOC pipeline on the 2-minute cadence \tess\ data was used to validate the candidate, however. According to the SPOC data validation report, the radius of the planet is $\sim 1.1 \, \mathrm{R}_\oplus$, which is consistent with our analysis. Consequently, the initial value derived from the QLP from a single sector was inaccurate, and the more detailed analysis performed in this study indicates that the radius is actually $\sim 1.0 \, \mathrm{R}_\oplus$.
This can also explain why we cannot identify the planetary signal in the RV data. The target was initially selected according to a mass estimate based on a planet radius of $R_{\rm p} = 1.89$ $\mathrm{R}_\oplus$, but since its actual radius is about half this value, the RV signal that it generates ($K \sim 0.5$\,m\,s$^{-1}$) is very difficult for current CARMENES capabilities and within the allocated observing time. 

To further assess the robustness of our derived mass uncertainties, we conducted a dedicated suite of simulations to quantify the number of observations required to constrain the planetary masses with a precision better than $15\%$ using CARMENES. The $15\%$ limit was chosen over a standard $5 \sigma \,(20\%)$ limit, based on the mutual mass-radius precision required to obtain reliable first-order inferences about the interior structure of the planets (\citealt{suissa_2018,caballero_2022,plotnykov2024}).

For each system, we generated synthetic RV time series, injected random noise (instrumental noise, stellar jitter, and an additional scaling factor to reproduce the observed RMS value), and fit the resulting synthetic curves. By increasing the sampling density, we found that about 80-100 observations are needed to determine the mass of TOI-1243\,b at a precision of 15\%. For TOI-4529\,b, the required number of observations is $\sim 190$-$210$, and for TOI-5388\,b, more than 300 observations are needed. A summary of this analysis is reported in the last column of Table~\ref{tab:tab:observing_params}.

\begin{figure}[ht]
        \includegraphics[width=0.5\textwidth]{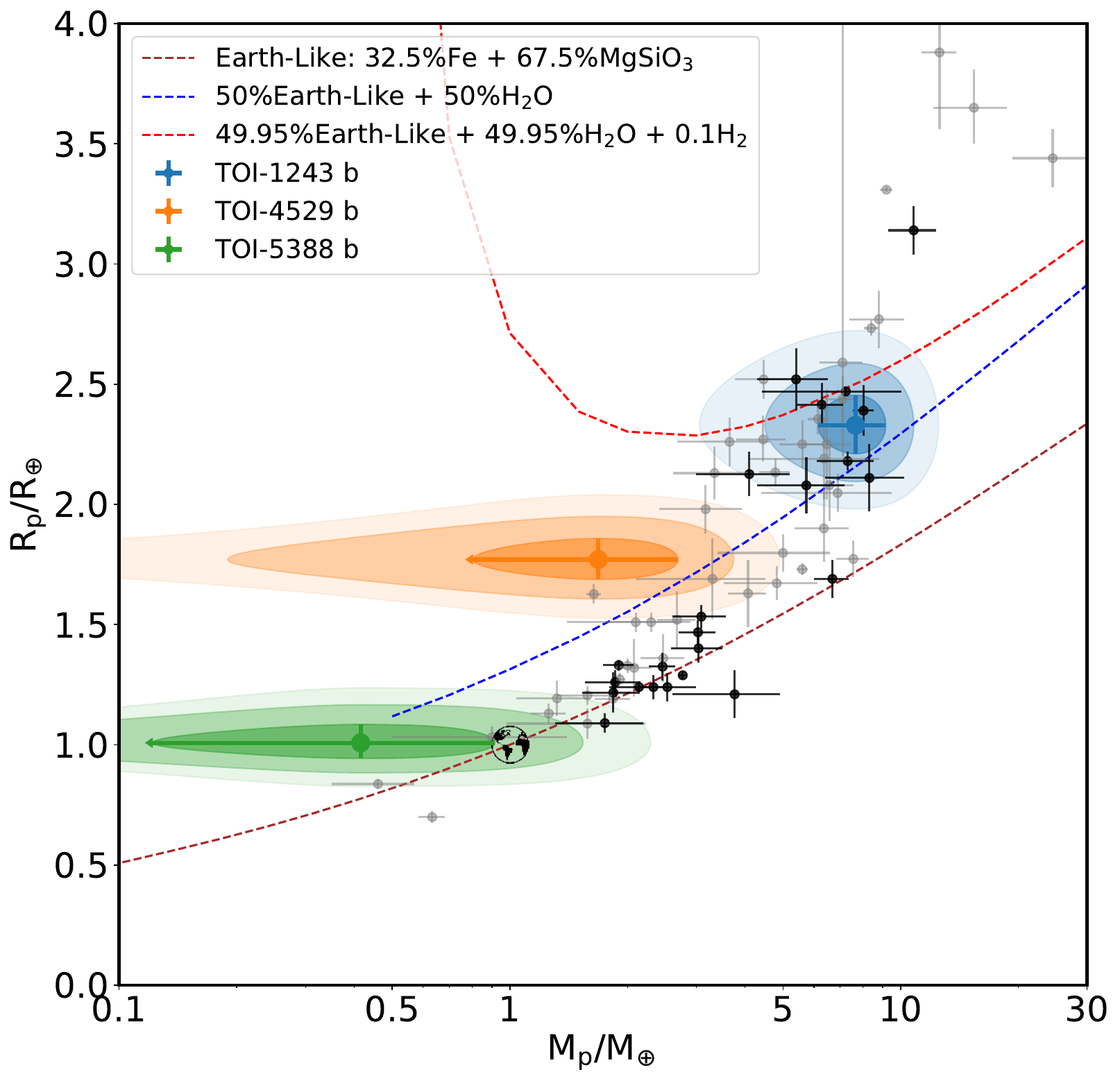}
        \caption{Mass-radius diagram for the three discussed planets in comparison with validated exoplanets around M dwarfs with masses of $M_{\rm p} < 30$\,M$_{\oplus}$ and $R_{\rm p} < 4$\,R$_{\oplus}$. The exoplanet data were downloaded from NASA Exoplanet Archive on 2025\,August\,2. The black dots highlight planets characterized with CARMENES. The diagram features several compositional lines as computed by \cite{planet_growth_models2019}. The $50\%$ Earth-like + $50\%$ H$_2$O and $49.95\%$ Earth-like + $49.95\%$ H$_2$O + 0.1 H$_2$ compositional lines are plotted for an equilibrium temperature of 500\,K, that is, the closest to the derived temperature of the exoplanets under analysis.}
\label{fig:mr}
\end{figure}

Our mass and radius determination for \plaa\ implies a bulk density of $ 3.3 \pm 0.9 \, \rm g\,cm^{-3}$, whereas for \plab\ and \plac\, we found  $\mathrm{3\sigma}$ density upper limits of $4.8$ and $8.9$ $\rm g\,cm^{-3}$, respectively.
Fig.~\ref{fig:mr} shows the mass-radius diagram for \plaa, \plab, and \plac in comparison with confirmed exoplanets around M dwarfs with masses of $M_{\rm p} < 30$ M$_{\oplus}$ and $R_{\rm p} < 4$ R$_{\oplus}$. This dataset includes 66 planets that fit our selection criteria. Twenty-four of these were published in papers led by the CARMENES team \footnote{From: \citealt{carm_luq_2019, carm_kemmer_2020, carm_blum_2020, carm_nowak_2020,Trifonov2021,carm_soto_2021, carm_blum_2021, carm_kossakowski_2021, carm_luque2022, carm_chaturvedi_2022,  carm_gonzalez_2022, carm_kemmer_2022, carm_espinoza_2022,carm_gonzales2023,carm_palle2023, carm_mallorquin2023, carm_murgas2024, carm_goffo2024}.} and are shown in black in Fig.~\ref{fig:mr}. The number above means that at least $\sim 36 \%$ of known sub-Neptunes around M-dwarf hosts have a determined mass based on CARMENES data, and several more have been characterized including CARMENES data together with additional RV datasets. This highlights its importance in the context of the characterization of small planets.

While \plac\ is most probably rocky based on its Earth-like radius and upper mass limit, the other two planets are situated in a highly degenerate region in the mass-radius plane (Fig.~\ref{fig:mr}), and more than one composition might therefore explain their mass, radius, and density. The mass of \plaa\ alone seems to be high enough for the planet to host a significant H-He envelope, although a water-world composition seems to be equally likely, and a pure rocky interior cannot be totally ruled out either. It is essential to reduce the uncertainty in the planet radius and mass for a better understanding of its interior composition. The uncertainty in the mass-radius space of \plab\  indicates that the planet is located closer to the water-world regime, and if this is the case, it would be one of the lowest-density planets of its size discovered around an M-dwarf star. A precise mass estimate is necessary before any conclusions are drawn, however. 

\subsection{Prospects for atmospheric characterization}
\label{ssec:atmospheric_characterization}

We calculated the transmission spectroscopy metric (TSM) and emission spectroscopy metric (ESM), as proposed by \cite{kempton2018}, to evaluate whether the three planets are suitable for atmospheric characterization with the \textit{JWST}. We obtained $\mathrm{TSM} = 50^{+15}_{-11}$ and $\mathrm{ESM} = 3.9^{+0.6}_{-0.5}$ for \plaa, $\mathrm{TSM} = 102^{+60}_{-31}$ and $\mathrm{ESM}=4.3^{+1.1}_{-1.0}$ for \plab, and $\mathrm{TSM} = 35^{+52}_{-17}$ and $\mathrm{ESM}=2.6^{+1.4}_{-0.7}$ for \plac\ (TSM and ESM values were calculated by sampling 100\,000 planet mass and radius values from their derived posterior distribution). All the $1\,\sigma$ TSM ranges for \plab\ include or are above the \cite{kempton2018} respective thresholds of 80 for small mini-Neptunes and 10 for terrestrial planets (\plac), which define the first quartile of the most feasible targets within their size classes, making them potentially appealing targets for future atmospheric characterization. More precise mass measurements are needed to meaningfully constrain their potential atmospheric scenarios, however. On the other hand, none of the three targets is particularly favorable for emission spectroscopy because their ESM values are all below the threshold recommended by \cite{kempton2018} for their planet size range (ESM = 7.5). 

We further explored the potential of \plaa, the only target in this study with a robust mass detection, for transmission spectroscopy with the \textit{JWST} through spectral simulations for a range of possible atmospheric scenarios. 
We modeled various H/He atmospheres with one and one hundred times scaled solar abundances, considering both clear and hazy conditions, as well as a pure water-vapor atmosphere. Synthetic transmission spectra were generated using \texttt{TauREx3} \citep{Waldmann2015,Al-refaie2021}. 
This set of reference models and the methods were analogous to those reported in previous papers for other planets (e.g., \citealp{Orell-Miquel2023,carm_palle2023,carm_goffo2024,Barkaoui2025}).

We used \texttt{ExoTETHyS} to simulate the corresponding \textit{JWST} spectra, as observed with the NIRISS-SOSS (0.6--2.8\,$\mu$m), NIRSpec-G395H (2.88--5.20\,$\mu$m), and MIRI-LRS (5--12\,$\mu$m) instrumental modes. We adopted a spectral resolution of $R\sim$100 for NIRISS-SOSS and NIRSpec-G395H and a constant bin size of 0.25\,$\mu$m for MIRI-LRS, following the recommendations of the \textit{JWST} Transiting Exoplanet Community Early Release Science team \citep{Carter2024,Powell2024}.

Fig.~\ref{fig:jwst_atmo} shows the synthetic transmission spectra for the atmospheric configurations above. The H/He model atmospheres exhibit strong H$_2$O and CH$_4$ absorption features up to a few hundred parts per million (ppm), depending on metallicity and haze, while the steam H$_2$O atmosphere has absorption features of $\lesssim$40\,ppm. The predicted error bars for a single transit observation are 38--158\,ppm (mean error 69\,ppm) for NIRISS-SOSS, 49--132\,ppm (mean error 74\,ppm) for NIRSpec-G395H, and 66--99\,ppm (mean error 78\,ppm) for MIRI-LRS. Our simulations suggest that one or a few transit observations are sufficient to detect an H/He atmosphere with up to $\sim$100$\times$ solar metallicity and some degree of haziness. More than five transits would be needed to detect a steam H$_2$O atmosphere.

\section{Conclusions}
\label{sec:conclusions}

We presented the confirmation and characterization of three new transiting sub-Neptunes, \plaa, \plab, and \plac\ around the nearby M-dwarf stars LSPM~J0902+7138, G~2--21, and Wolf~346, respectively.
We identified them through \tess\ transit observations and confirmed them using ground-based photometric observations and RV measurements obtained with the CARMENES instrument. These planets exhibit distinct properties that highlight the diversity of sub-Neptunes around low-mass stars.

\plaa\ has an orbital period of 4.66\,d. Its derived mass of $M_{\rm p} = 7.7\pm1.5 \, {\rm M}_{\oplus}$ and radius of $R_{\rm p} = 2.33 \pm 0.12 \, {\rm R}_{\oplus}$ imply a density of $ 3.3 \pm 0.9$\,g\,cm$^{-3}$, suggesting that it might be a small gaseous planet or a water world. Our atmospheric modeling shows that a few (even one) transit observations with JWST will allow for the detection of a H/He atmosphere, if present, while the detection of a steam H$_2$O atmosphere would require the observation of at least five transits. 

\plab\ has an orbital period of 5.88 days, a radius of $R_{\rm p} = 1.77^{+0.09}_{-0.08} \, {\rm R}_{\oplus}$, and a $3 \sigma$ upper mass limit of $M_{\rm p} < 4.9 \, {\rm M}_{\oplus}$, resulting in an upper density limit of $4.8$\,g\,cm$^{-3}$, compatible with a water-world composition. 

\plac\ is the smallest of the planets we analyzed. It has a short orbital period of 2.59 days, a radius of $R_{\rm p} = 0.99^{+0.07}_{-0.06} \, {\rm R}_{\oplus}$, and a $\mathrm{3 \sigma}$ upper mass limit of 2.2\,M$_{\oplus}$. Its radius suggests a rocky nature. 

Following an additional refinement of their masses, \plab\ and \plac\ might both be interesting for transmission spectroscopy studies because their TSM values are high.

\section*{Data availability}\label{sec:data_availability}
All transit LCs used in this work are available at ExoFOP. The CARMENES RVs and activity indicators are available in electronic form at the Centre de Donn\'ees astronomiques de Strasbourg via anonymous ftp to cdsarc.u-strasbg.fr (130.79.128.5) or via \url{http://cdsweb.u-strasbg.fr/cgi-bin/qcat?J/A+A/*/*}. 

\begin{acknowledgements}

We thank the anonymous referee for the comments that helped improving the quality of this paper. CARMENES is an instrument at the Centro Astron\'omico Hispano en Andaluc\'ia (CAHA) at Calar Alto (Almer\'{\i}a, Spain), operated jointly by the Junta de Andaluc\'ia and the Instituto de Astrof\'isica de Andaluc\'ia (CSIC). CARMENES was funded by the Max-Planck-Gesellschaft (MPG), the Consejo Superior de Investigaciones Cient\'{\i}ficas (CSIC), the Ministerio de Econom\'ia y Competitividad (MINECO) and the European Regional Development Fund (ERDF) through projects FICTS-2011-02, ICTS-2017-07-CAHA-4, and CAHA16-CE-3978, and the members of the CARMENES Consortium (Max-Planck-Institut f\"ur Astronomie, Instituto de Astrof\'{\i}sica de Andaluc\'{\i}a, Landessternwarte K\"onigstuhl, Institut de Ci\`encies de l'Espai, Institut f\"ur Astrophysik G\"ottingen, Universidad Complutense de Madrid, Th\"uringer Landessternwarte Tautenburg, Instituto de Astrof\'{\i}sica de Canarias, Hamburger Sternwarte, Centro de Astrobiolog\'{\i}a and Centro Astron\'omico Hispano-Alem\'an), with additional contributions by the MINECO, the Deutsche Forschungsgemeinschaft (DFG) through the Major Research Instrumentation Programme and Research Unit FOR2544 ``Blue Planets around Red Stars'', the Klaus Tschira Stiftung, the states of Baden-W\"urttemberg and Niedersachsen, and by the Junta de Andaluc\'{\i}a. Funding for the \tess\ mission is provided by NASA's Science Mission Directorate. This paper made use of data collected by the TESS mission and are publicly available from the Mikulski Archive for Space Telescopes (MAST) operated by the Space Telescope Science Institute (STScI). We acknowledge the use of public TESS data from pipelines at the TESS Science Office and at the TESS Science Processing Operations Center. Resources supporting this work were provided by the NASA High-End Computing (HEC) Program through the NASA Advanced Supercomputing (NAS) Division at Ames Research Center for the production of the SPOC data products. We acknowledge support from the \tess\ mission via subaward s3449 from MIT. This work makes use of observations from the LCOGT network. Part of the LCOGT telescope time was granted by NOIRLab through the Mid-Scale Innovations Program (MSIP). MSIP is funded by NSF. This article is based on observations made with the MuSCAT2 instrument, developed by ABC, at Telescopio Carlos S\'anchez operated on the island of Tenerife by the IAC in the Spanish Observatorio del Teide, and with the MuSCAT3 instrument, developed by the Astrobiology Center and under financial supports by JSPS KAKENHI (JP18H05439) and JST PRESTO (JPMJPR1775), at Faulkes Telescope North on Maui, HI, operated by the Las Cumbres Observatory. The Joan Or\'o Telescope (TJO) at the Observatori del Montsec (OdM) is owned by the Catalan Government and operated by the Institute of Space Studies of Catalonia (IEEC). This research has made use of the Exoplanet Follow-up Observation Program (ExoFOP; DOI: 10.26134/ExoFOP5) website, which is operated by the California Institute of Technology, under contract with the National Aeronautics and Space Administration under the Exoplanet Exploration Program. We acknowledge financial support from the Agencia Estatal de Investigaci\'on (AEI/10.13039/501100011033) of the Ministerio de Ciencia e Innovaci\'on and the ERDF ``A way of making Europe'' through projects PID2023-150468NB-I00, PID2022-137241NB-C4[1:4], PID2021-125627OB-C31, PRE2020-093107 (FPI-SO), RYC2022-037854-I, the internal project 20235AT003 associated to RYC2021-031640-I, and the Centre of Excellence ``Severo Ochoa'' and ``Mar\'ia de Maeztu'' awards to Instituto de Astrof\'isica de Andaluc\'ia (CEX2021-001131-S) and Institut de Ci\`encies de l'Espai (CEX2020-001058-M). This work is partly supported by JSPS KAKENHI Grant Numbers JP24H00017, JP24K00689, JSPS Bilateral Program Number JPJSBP120249910, JSPS Grant-in-Aid for JSPS Fellows Grant Numbers JP25KJ1036, JP25KJ0091, and JST SPRING, Grant Number JPMJSP2108.BBO and LT-O acknowledge support from The Israel Ministry of Innovation, Science, and Technology through Grant number 0008107, and by The Israel Science Foundation through grant No. 1404/22. This work made use of the following softwares: \texttt{AstroImageJ} \citep{Collins:2017}, \texttt{astropy} \citep{astropy2022}, \texttt{batman} \citep{batman2015}, \texttt{ExoTETHyS} \citep{Morello2020a}, \texttt{Pylightcurve}\citep{pylightcurve},  \texttt{juliet} \citep{juliet_2019}, \texttt{numpy} \citep{numpy2020}, \texttt{scipy} \citep{scipy2020}, \texttt{matplotlib} \citep{matplotlib2007}.

\end{acknowledgements}

\bibliographystyle{aa}
\bibliography{biblio}

\begin{appendix}
\onecolumn
\section{Photometric validation}

We report here the example plots of the photometric validation of the three candidates as described in Sect.~\ref{sec:photometric_vetting}.

\begin{figure*}[h!]
    \centering
    \includegraphics[width=0.54\textwidth]{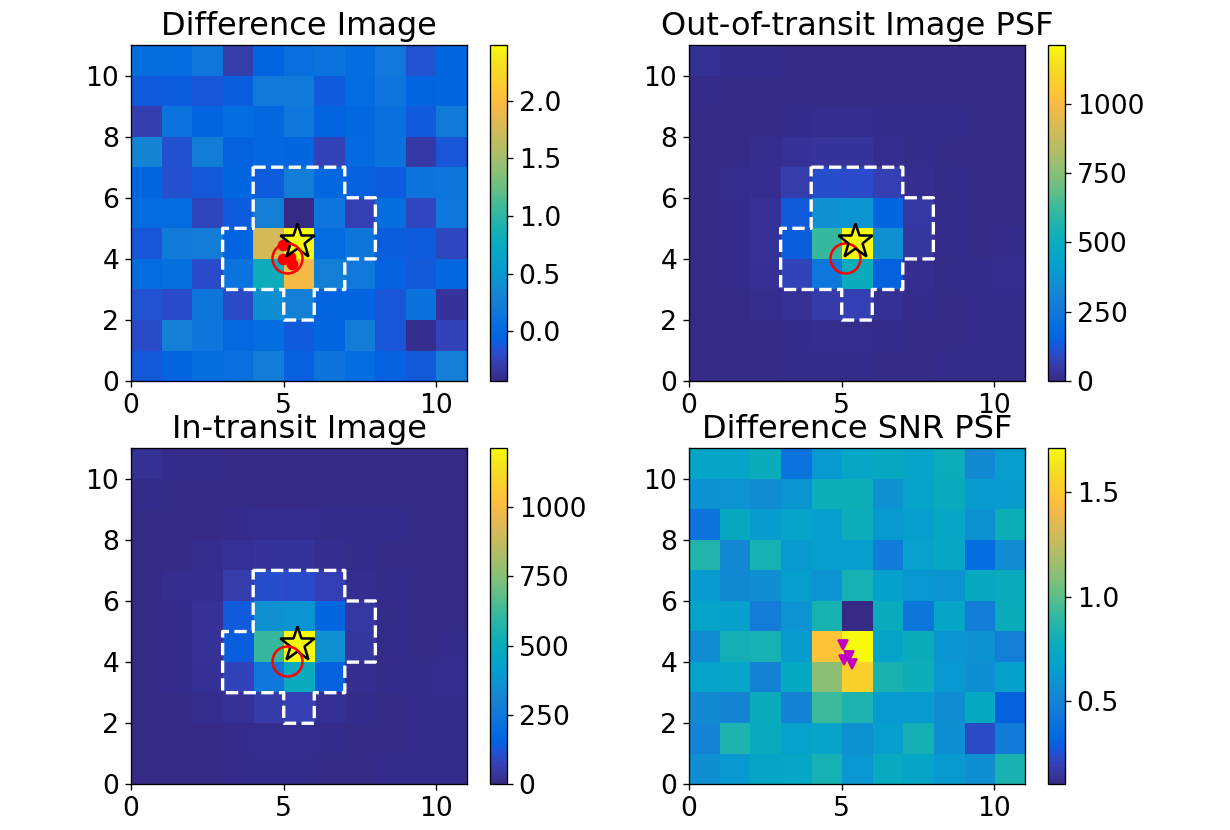}
    \centering
    \includegraphics[width=0.54\textwidth]{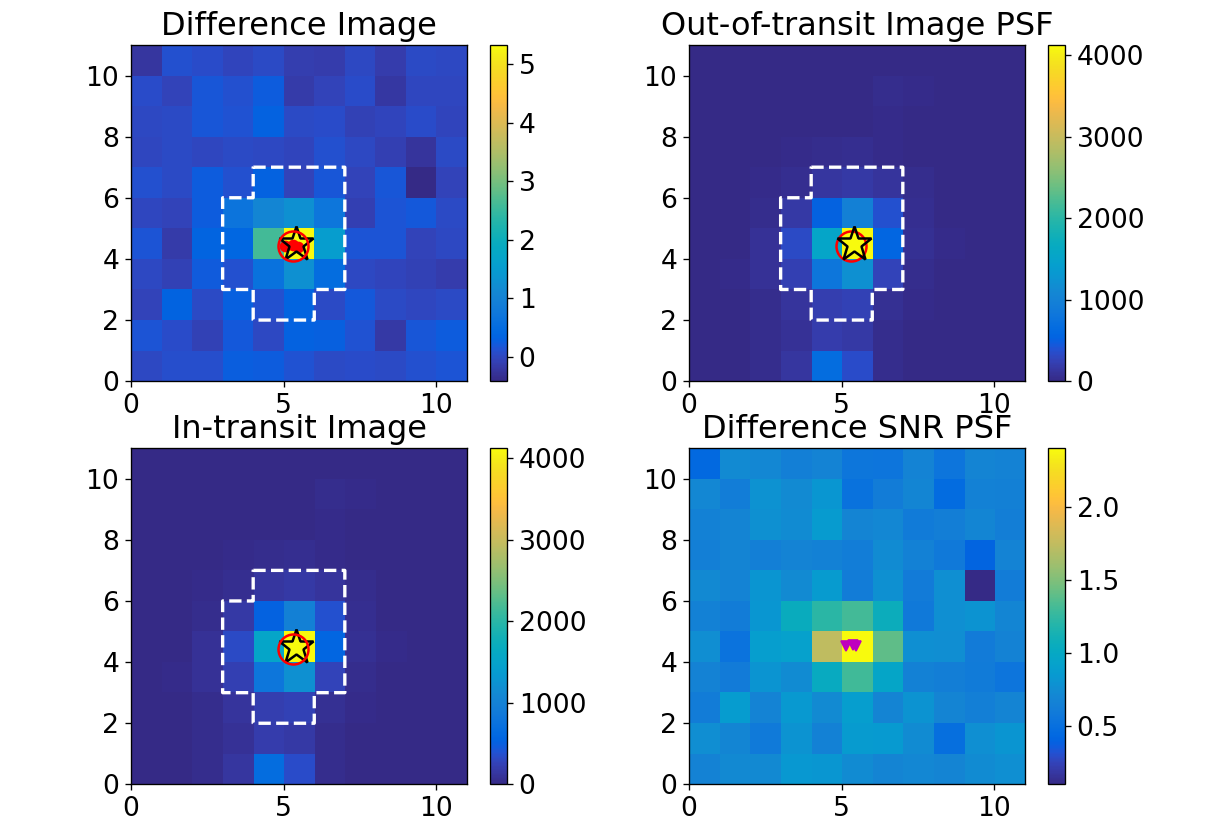}
    \centering
    \includegraphics[width=0.54\textwidth]{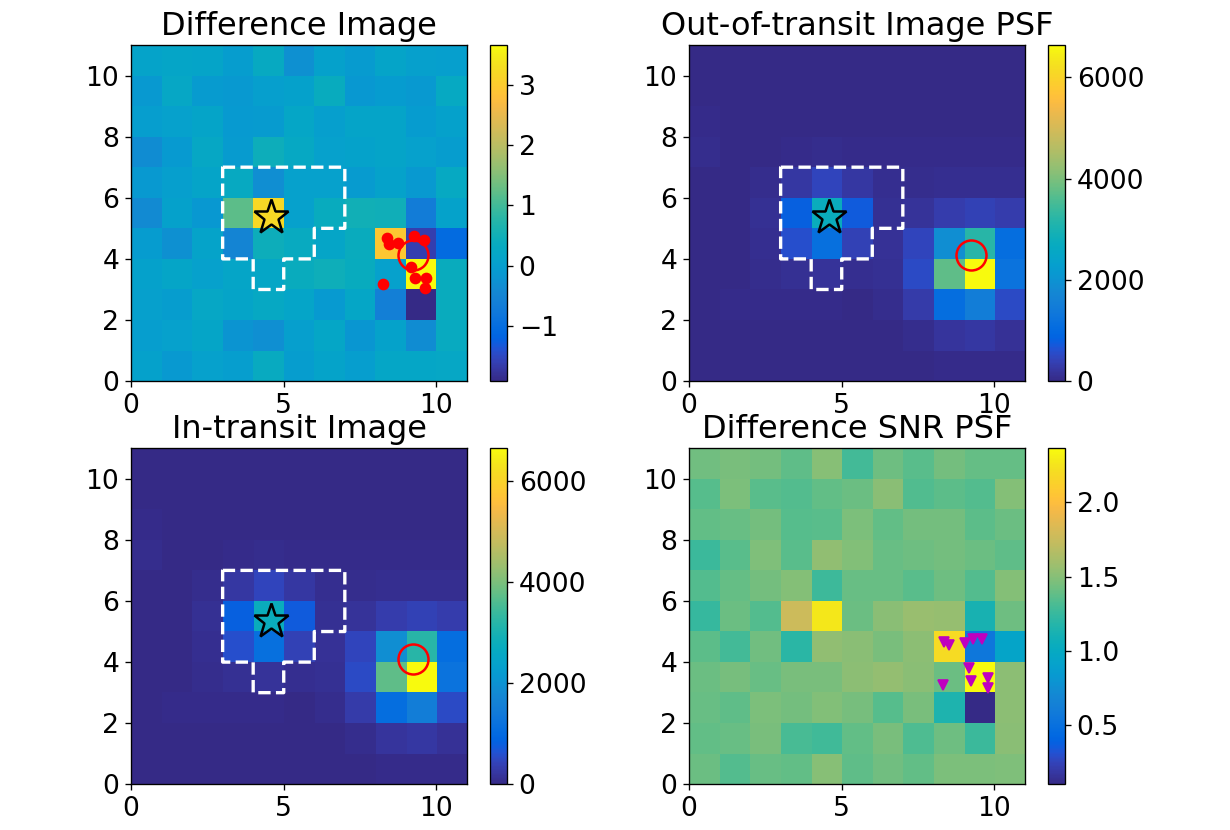}
    \caption{Output from the \texttt{centroids} module for \plaa\ in \tess\ Sector 60 (top), for \plab\ in \tess\ Sector 42 (middle), and for \plac\ in \tess\ Sector 21 (bottom). The dashed white lines outline the aperture mask for LC extraction. A star symbol indicates the cataloged position of the target, while a purple triangle shows the average out-of-transit photocenter. Individual photocenters are marked by small red dots, and the large red circle denotes the overall difference image photocenter. The panels show: upper left - difference image; upper right - average out-of-transit image; lower left - average in-transit image; lower right - S/N of the mean difference image. A color bar indicates the number of electrons/sec for each case mentioned. The difference image reveals a centroid offset without artifacts slightly contaminated by the presence of a nearby source.}
    \label{fig:1243_centroids}
\end{figure*}

\FloatBarrier 
\begin{figure*}[h!]
    \centering
    \includegraphics[width=0.9\textwidth]{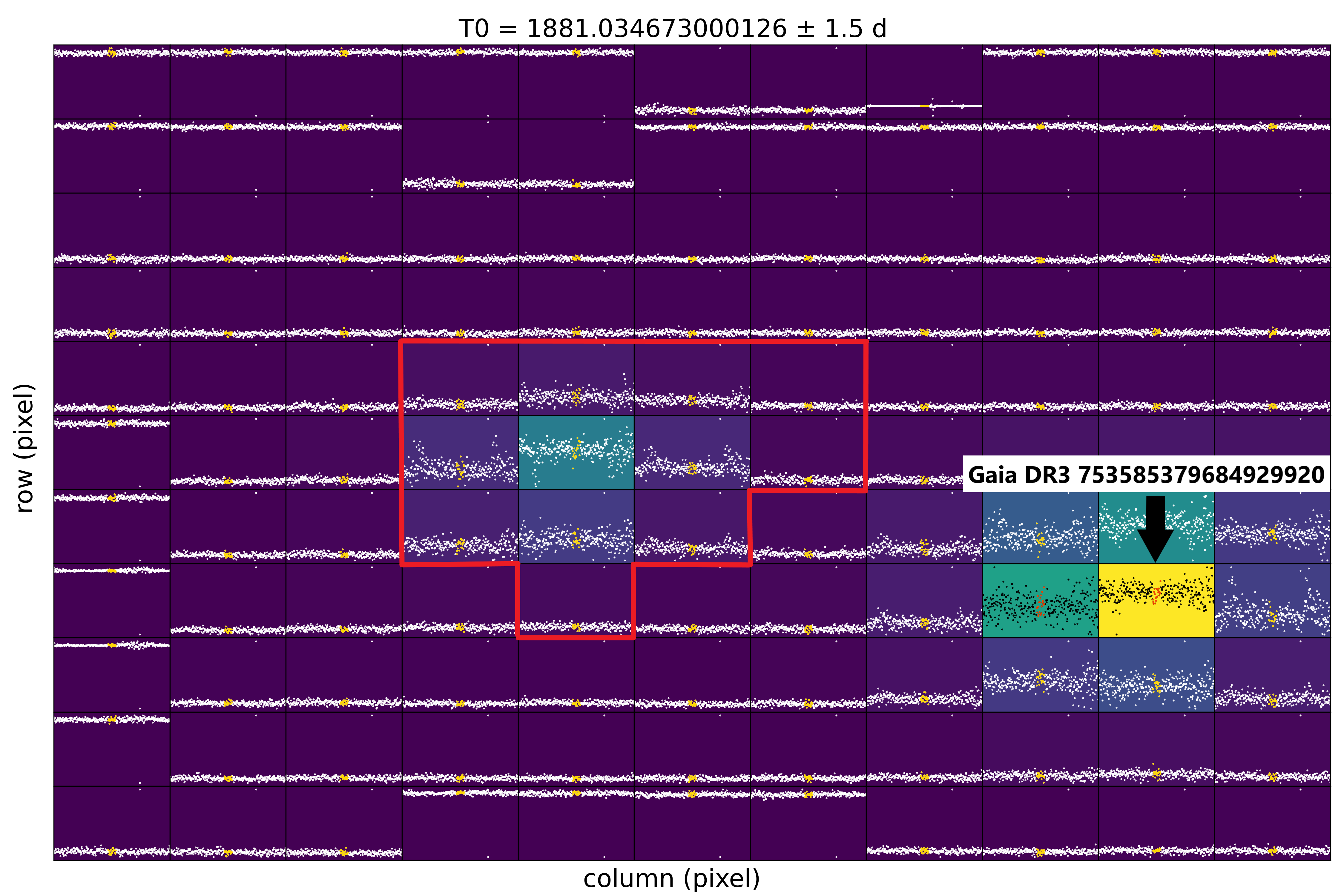}
    \caption{The PLL analysis of the fifth detected transit of \plac\ at $1881.03$ TBJD. Each square represents the flux of a pixel, with respect to time, and the color scale represent the average flux value. Yellow dots represent the expected transit time. The red contour represents the aperture mask for LC extraction.}
    \label{fig:PLL_TOI5388}
\end{figure*}

\FloatBarrier

\section{Observations Summary}
This section provides a summary of photometric and spectroscopic observations used in this work.
\begin{table*}[h!]
    \caption{Summary the \tess\ observations.}
    \centering
    \begin{tabular}{cccccc}
        \hline
        \hline
        TOI ID & TIC ID & Obs. start date & Obs. end date & \tess\ Sectors & Number of transits \\
        \hline
        \splaa\ & 219698776 & 2019\,Jul\,18    & 2024\,Jan\,30 & 14, 20, 40, 47, 53, 54, 74 & 33 \\
        \splab\ & 384888319 & 2021\,Aug\,20  & 2023\,Oct\,16 & 42, 43, 70                 & 9\\
        \splac\ & 407591297 & 2020\,Jan\,21 & 2022\,Feb\,26 & 21, 48                    & 18 \\
        \hline
    \end{tabular}
    \label{tab:tess_observations}
\end{table*}

\begin{table*}[h!]
    \caption{Summary the CARMENES observations.}
    \centering
    \small
    \begin{tabular}{cccccccc}
        \hline
        \hline
        Name & Start date & End date & N &  Median internal precision & RMS & S/N (median) &$\mathrm{N\,({M_{unc}<15\%}})^{(a)}$\\
        \hline
        \splaa\ & 2022\,Sep\,18 & 2024\,Jun\,26 & 39 &  2.82 $\mathrm{m\,s^{-1}}$ & 5.09  $\mathrm{m\,s^{-1}}$ & 245 &  80-100\\
        \splab\ & 2022\,Jul\,10 & 2024\,Jan\,31 & 62 &  1.91  $\mathrm{m\,s^{-1}}$ & 4.82  $\mathrm{m\,s^{-1}}$& 457 &190-210\\
        \splac\ & 2022\,Apr\, 02 &  2024\,Jun\,23 & 42 & 2.2  $\mathrm{m\,s^{-1}}$ & 4.18  $\mathrm{m\,s^{-1}}$ & 333 & >300\\
        \hline
    \end{tabular}
    \tablefoot{
    Observations were collected under the CARMENES programs 22A-3.5-005, 22B-3.5-006, 23A-3.5-006, 23B-3.5-008, 24A-3.5-006. $^{(a)}$ Predicted number of required CARMENES RV observations to achieve a mass precision lower than $15 \%$.}
    \label{tab:tab:observing_params}
\end{table*}

\twocolumn

\begin{table}[h!]
    \caption{Summary of ground-based transit  observations, as described in Sect.~\ref{ssec:transit_photometry}.}
    \centering
    \begin{tabular}{cccc}
        \hline
        \hline
        Planet Name & Obs. Date & Instrument & Filter  \\
        \hline
        \plaa\ & 2021\,Feb\,03 & 1m/McD & $i'$ \\
        \plaa\ & 2021\,Feb\,12 & 2m/MuSCAT3 & $g'$, $r'$, $i'$, $z_s$ \\
        $^*$\plab$^p$ & 2021\,Nov\,12 & 1m/SAINT-EX & $z_s'$  \\
        $^*$\plab$^p$ & 2022\,Jan\,21 & 1.5m/MuSCAT2 & $g'$, $r'$, $z_s$  \\
        $^*$\plab$^p$ & 2022\,Oct\,30 & 1.5m/MuSCAT2 & $g'$, $r'$, $i'$, $z_s$ \\
        $^*$\plab\ & 2022\,Dec\,17 & 1m/CTIO & $z_s$ \\
        $^*$\plab\ & 2023\,Sep\,13 & 1m/SSO & $z_s$ \\
        \plac\ & 2022\,Mar\,26 & 1m/McD & $i'$\\
        \plac\ & 2022\,Apr\,08 & 1m/CTIO & $i'$\\
        \plac\ & 2022\,Apr\,15 & 1m/TEID & $z_s$\\
        \plac\ & 2022\,Apr\,26 & 1m/SAINT-EX & $r'$  \\
        $^*$\plac\ & 2023\,Mar\,13 & 1.5m/MuSCAT2 & $g'$, $r'$, $z_s$ \\
        $^*$\plac\ & 2024\,Apr\,26 & 1.5m/MuSCAT2 & $g'$, $r'$, $i'$, $z_s$\\
        \hline
    \end{tabular}
\tablefoot{{p} Partial transits. $^*$ The transit LC was omitted from the global fit, as it did not pass the selection criteria.}
    \label{tab:ground_based_transits}
\end{table}

\begin{table}[h!]
    \caption{Summary of the ground-based long-term photometric follow-up observations, as described in Sect.~\ref{ssec:long_term_photometry}}
    \centering
    \begin{tabular}{cccccccc}
        \hline
        \hline
        Star Name & Obs. Start & Obs. End & Instrument & Filter \\
        \hline
        \splaa\ & 2024\,Mar & 2024\,Jun & 1m/McD & $V$\\
        \splaa\ & 2024\,Apr & 2024\,May & 0.4m/LCO & $V$\\
        \splaa\ & 2024\,Mar & 2025\,Jan & 0.8m/TJO  & $R$\\
        \splaa\ & 2024\,Mar & 2025\,Jan & 0.4m/e-EYE  &  $V$, $R$  \\

        \splab\ & 2024\,Sep & 2024\,Dec & 1m/McD & $B$\\
        \splab\ & 2024\,Aug & 2024\,Dec & 0.4m/LCO & $V$\\
        \splab\ & 2023\,Mar & 2025\,Jan & 0.8m/TJO  & $R$  \\
        \splab\ & 2024\,May & 2025\,Jan & 0.4m/e-EYE  & $V$, $R$  \\

        \splac\ & 2024\,Mar & 2024\,Jun & 1m/McD & $B$\\
        \splac\ & 2024\,Apr & 2024\,Jun & 0.4m/LCO & $V$\\
        \splac\ & 2024\,Apr & 2025\,Jan & 0.8m/TJO  & $R$   \\
        \splab\ & 2024\,Mar & 2025\,Jan & 0.4m/e-EYE  & $V$, $R$  \\

        \hline
    \end{tabular}
    \label{tab:ground_based_longterm}
\end{table}

\section{Periodogram analysis}
\label{app:periodograms}

GLS periodograms of the CARMENES RV data and the stellar activity indicators for the three planets under investigation. For \plab, we also report the periodograms of the long-term ground-based photometry and \tess\ data. 
\tess\ and long-term ground-based periodograms for \plaa\ and \plac\ 
are not reported here, as they do not exhibit any statistically significant peak.

\begin{figure}[h!]
    \centering    \includegraphics[width=0.5\textwidth]{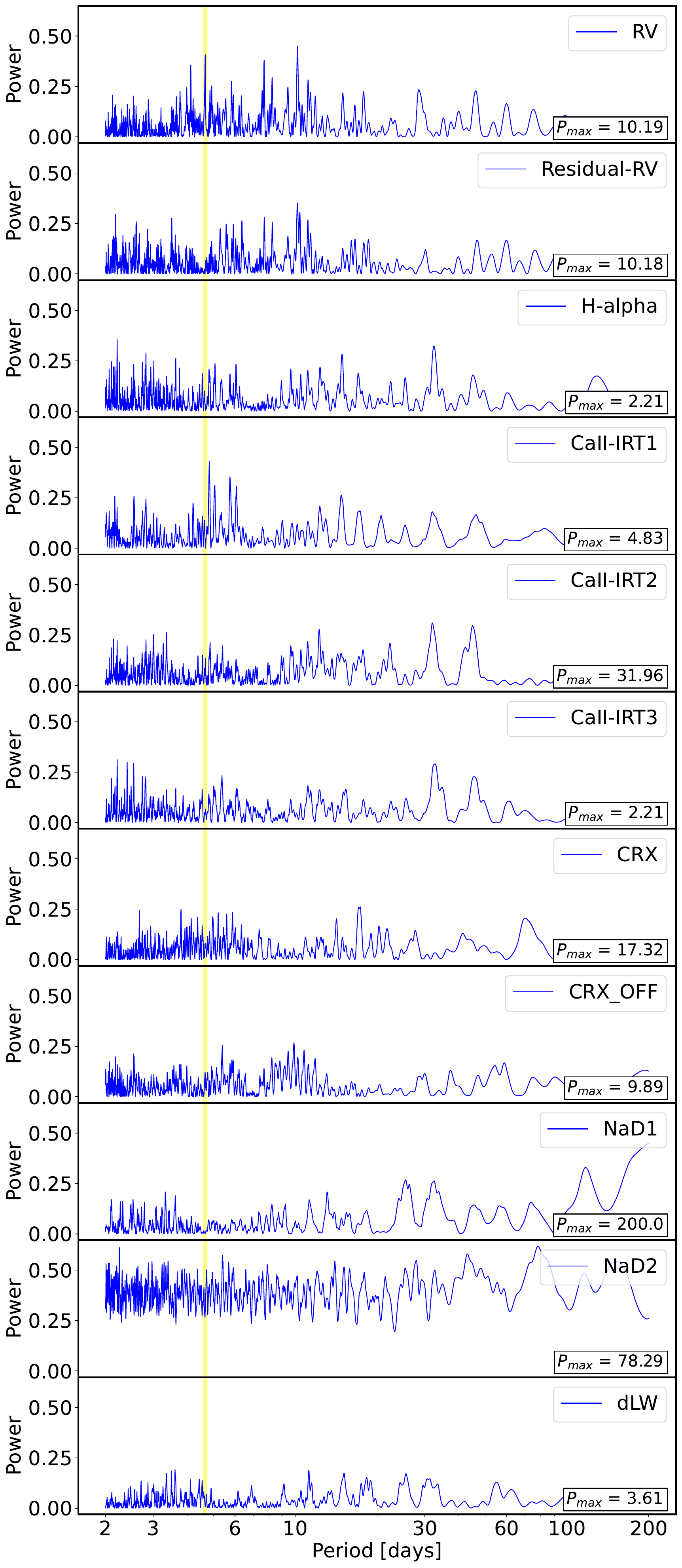}
        \caption{Periodograms of the spectroscopic data of \plaa. The first and second panels show the RV time series, and the residuals of the RV after the best-fit model. The additional panels show the periodograms of activity indicators. The yellow vertical line represents the expected 4.66~d transit period of the planet from the photometric analysis.}
            \label{fig:per1243}
\end{figure}

\begin{figure}[h!]
    \centering
        \includegraphics[width=0.5\textwidth]{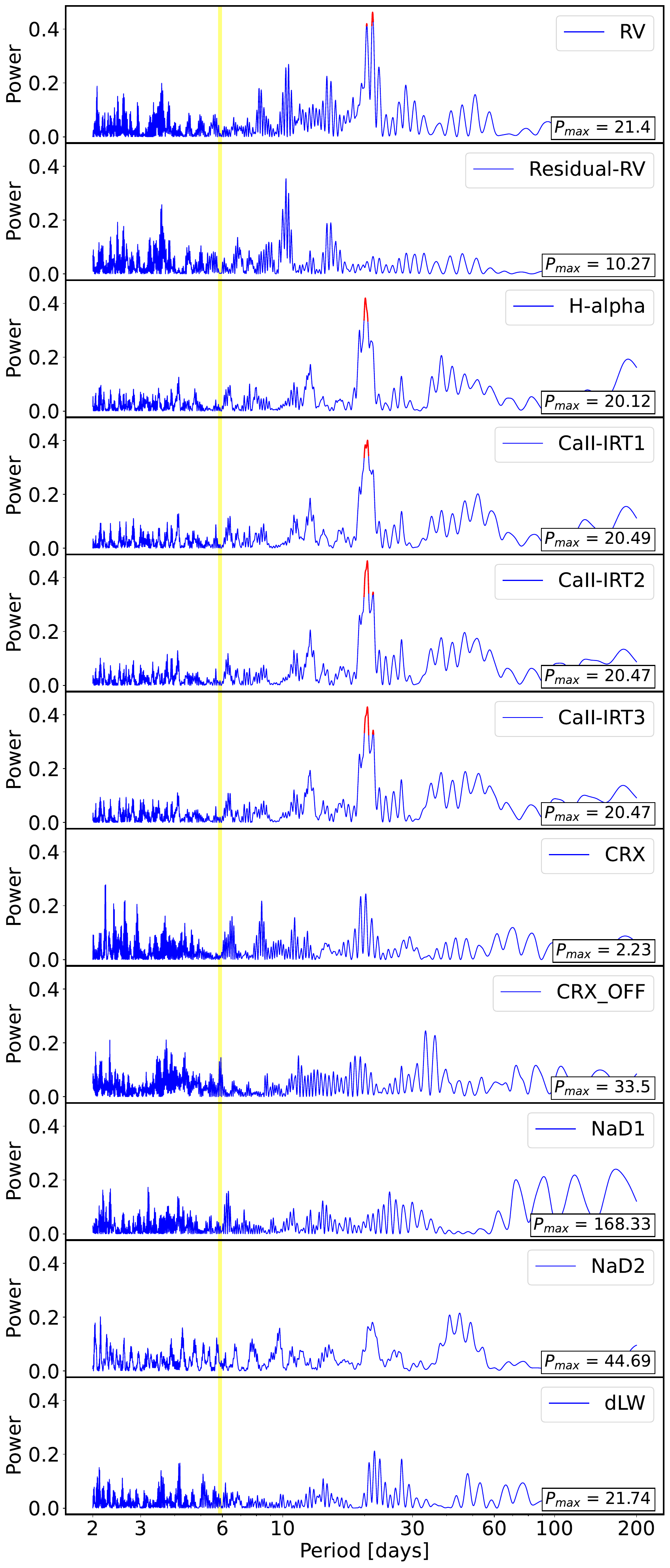}
        \caption{Same as Fig.~ \ref{fig:per1243}, but for \plab. The red marked power represent periods with FAP<$0.13\%$.The yellow vertical line represents the expected 5.38~d transit period of the planet from the photometric analysis.}
    \label{fig:per4529}
\end{figure}

\begin{figure}[h!]
    \centering
        \includegraphics[width=0.5\textwidth]{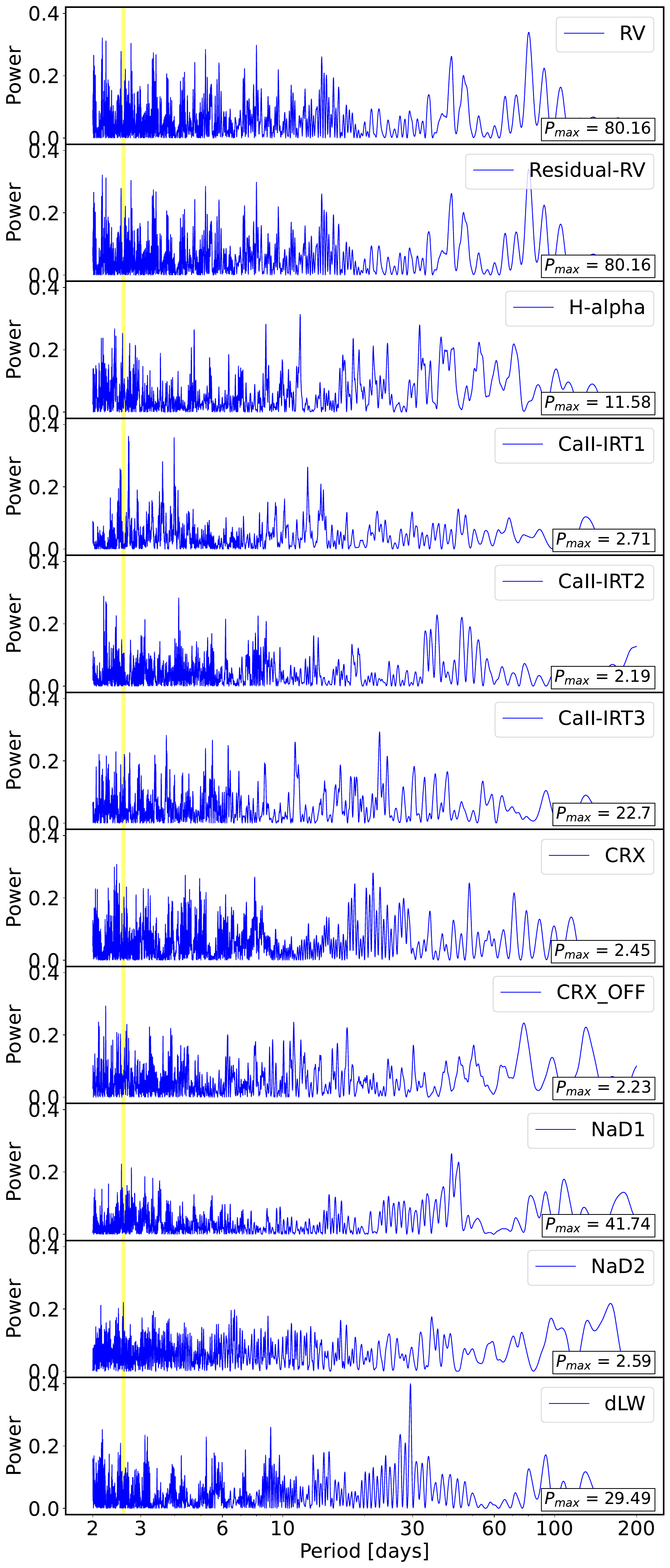}
        \caption{Same as Fig.~\ref{fig:per1243}, but for \plac. The yellow vertical line represents the expected 5.38~d transit period of the planet from the photometric analysis.}  
    \label{fig:per5388}
\end{figure}

\onecolumn
\begin{figure*}[t]
    \centering
        \includegraphics[width=\textwidth]{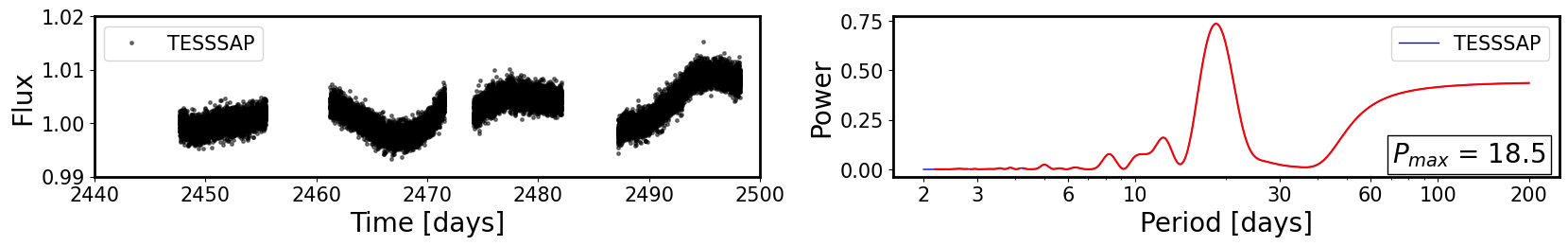}
    \par
    \centering
        \includegraphics[width=\textwidth]{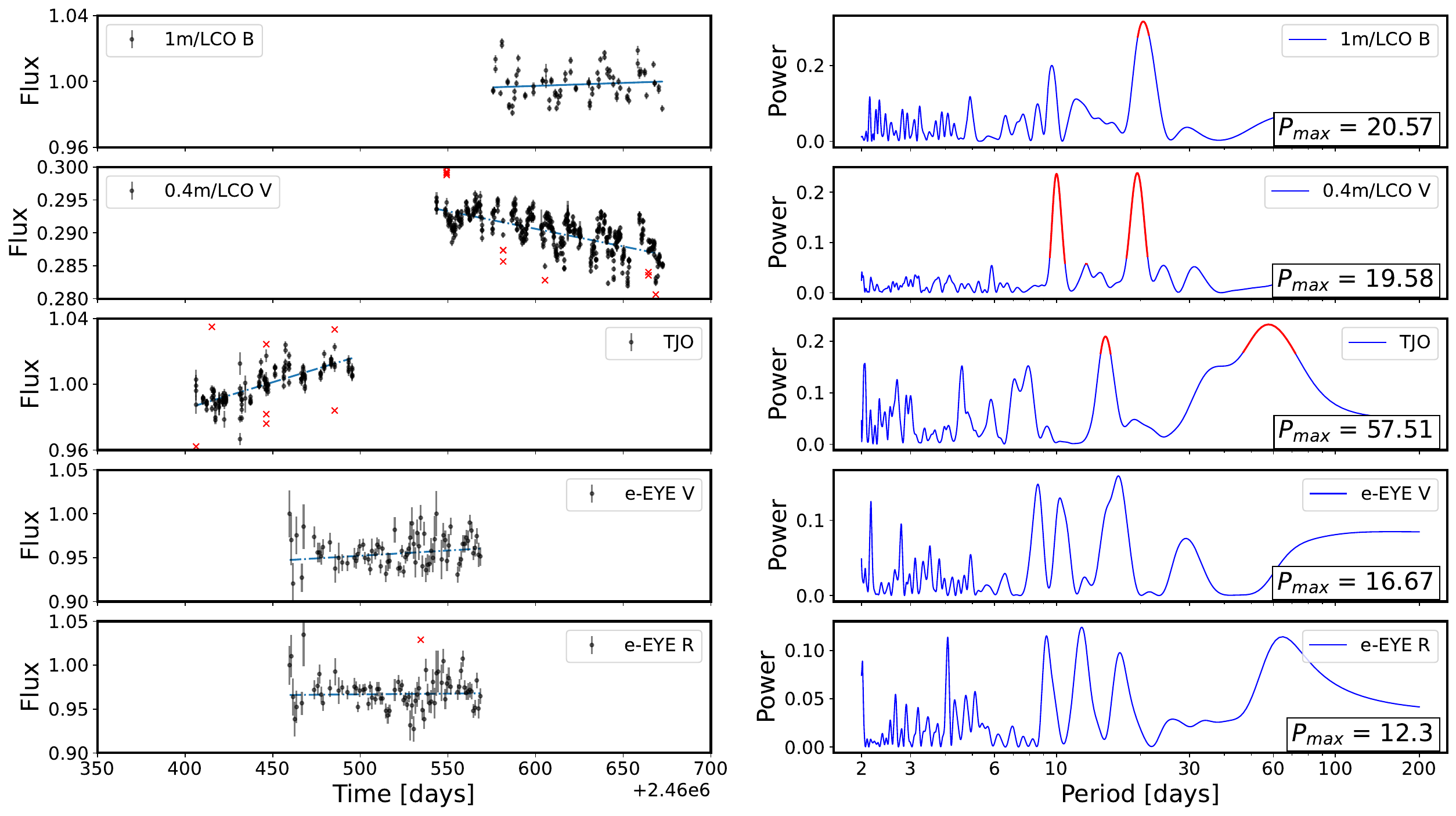}
    \caption{Top: \tess\ SAP data for \splab\ of the consecutive Sectors 42, 43. Sector 70 is not included in this analysis to avoid the effect of the large-gap window function that appears in the periodogram. Bottom: Ground-based long-term photometric data for \splab. The left panels show the relative flux. Red crosses mark are outliers not used in the analysis. The right panels show the GLS periodograms of the flux, with the horizontal dashed line showing the $3 \sigma$ (0.13 \%) FAP, and the vertical red line showing the period of the most significant peak. 
}
    \label{fig:4529longterm}
\end{figure*}

\section{Ground-based LCs}
\label{appendix:ground_based_lcs}
We report here the fit of all the ground-based follow-up observations of the three planets.

\begin{figure*}[h!]
    \centering
        \includegraphics[width=0.9\textwidth]{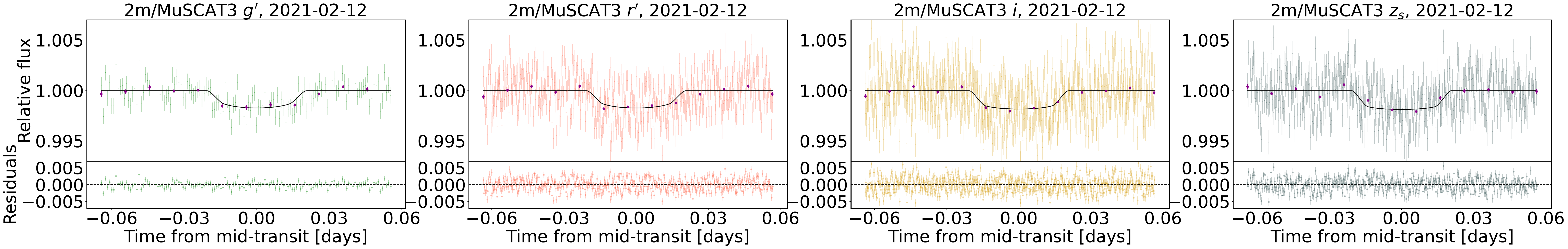}
    \par
    \centering
        \includegraphics[width=0.225\textwidth]{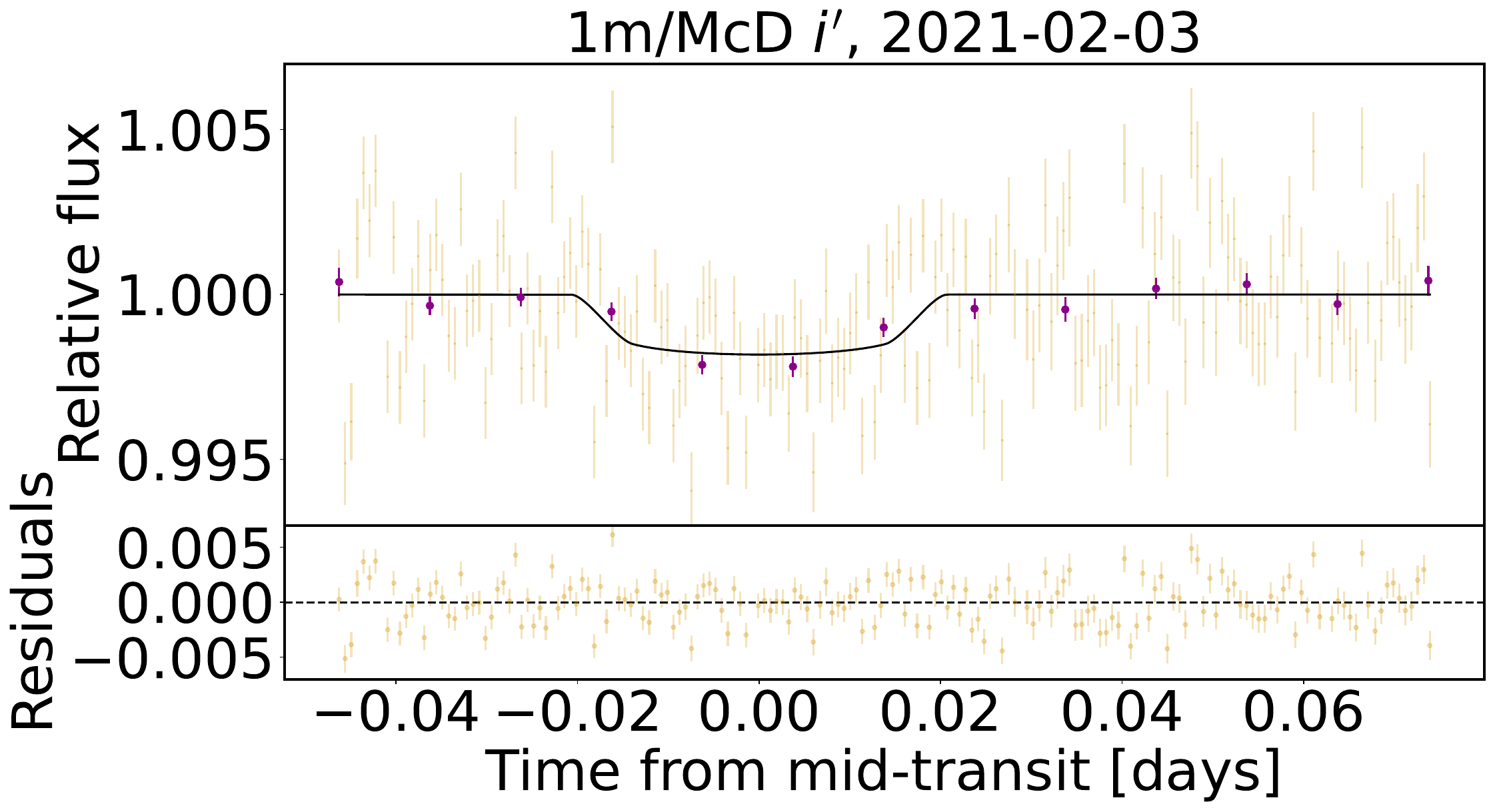}
    \caption{Ground-based follow-up observations of \plaa. The purple points show the binned LCs, and the black lines shows the best-fit model. }
    \label{fig:g1243_1}
\end{figure*}

\begin{figure*}[h!]
    \centering
        \includegraphics[width=0.9\textwidth]{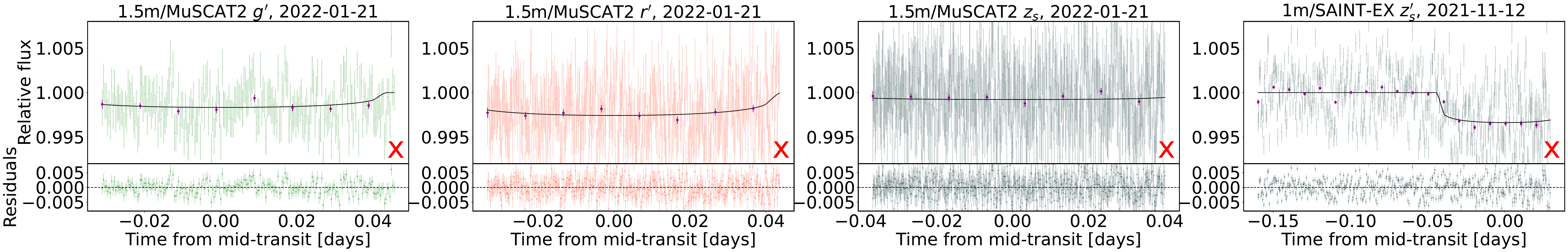}
    \par
    \centering
        \includegraphics[width=0.9\textwidth]{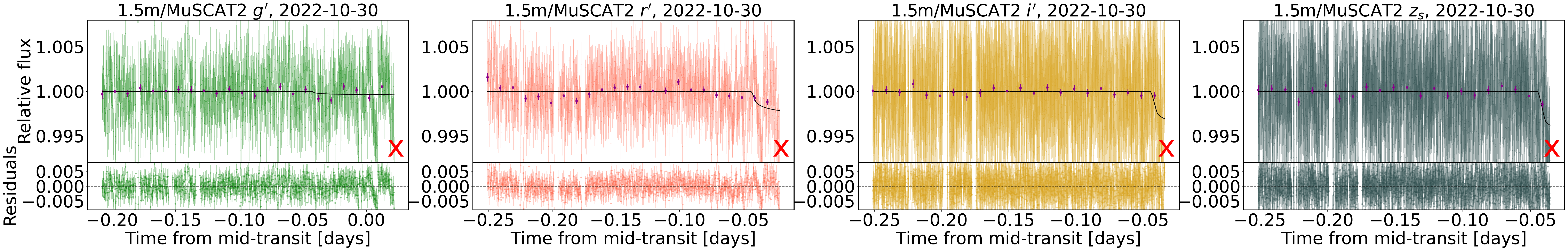}

    \par
    \centering
        \includegraphics[width=0.45\textwidth]{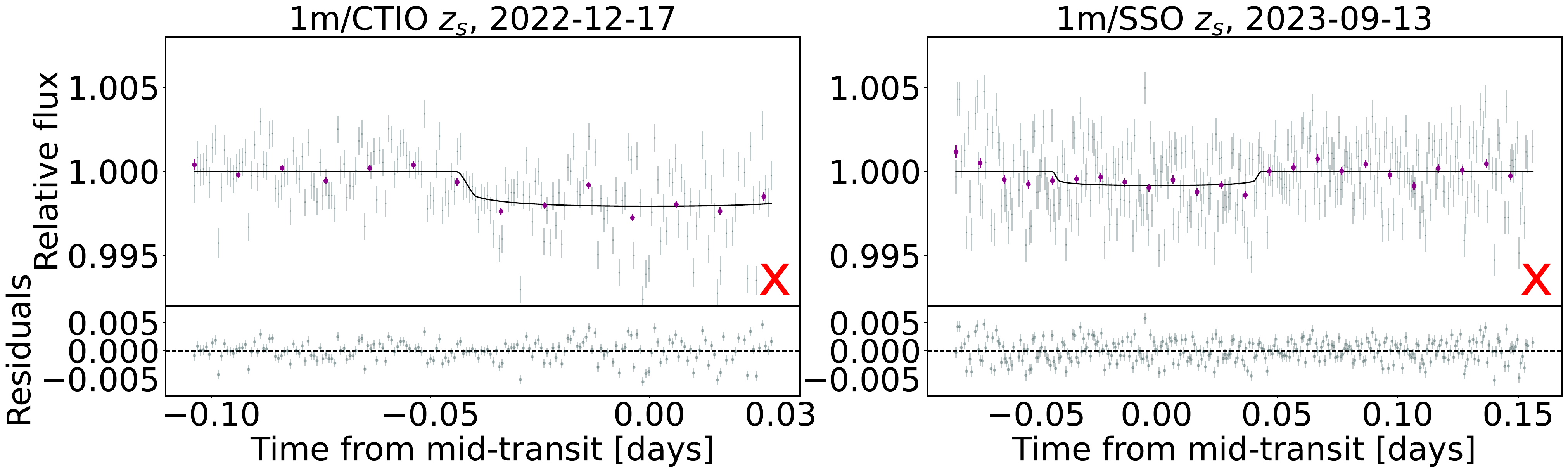}
    \caption{Same as Fig.~\ref{fig:g1243_1}, but for \plab. More that $7 \sigma$ outliers have been removed for better visualization. Red "X" tick indicates that the specific transit was not used in the global fit.}
    \label{fig:g4529}
\end{figure*}

\begin{figure*}[h!]
    \centering
       \includegraphics[width=0.675\textwidth]{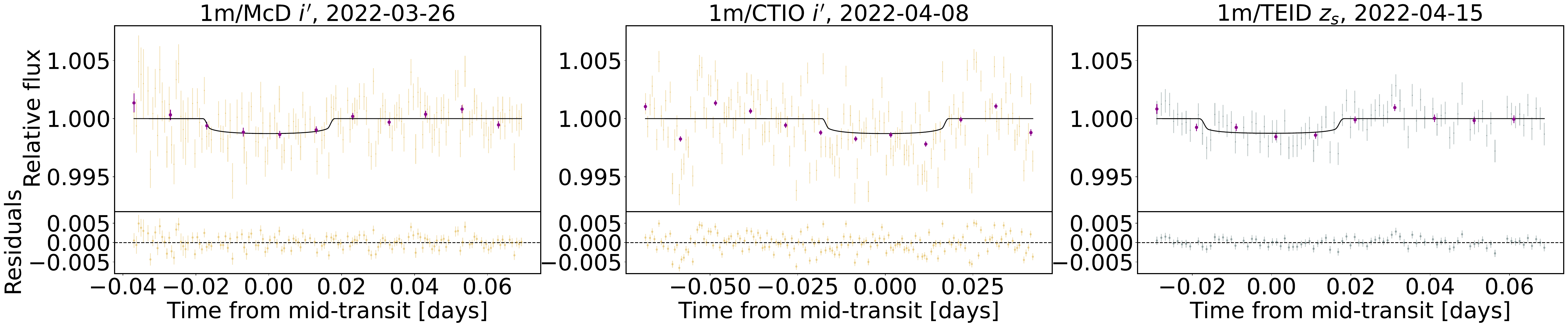}
    \par
    \centering
         \includegraphics[width=0.9\textwidth]{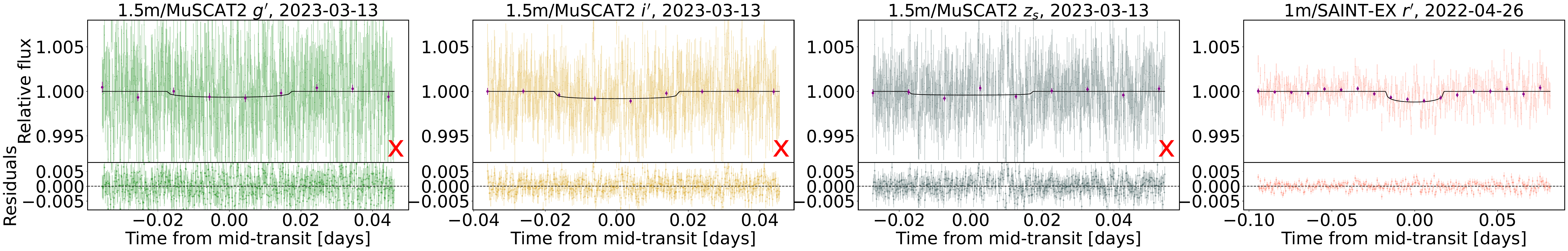}

    \par
    \centering
        \includegraphics[width=0.9\textwidth]{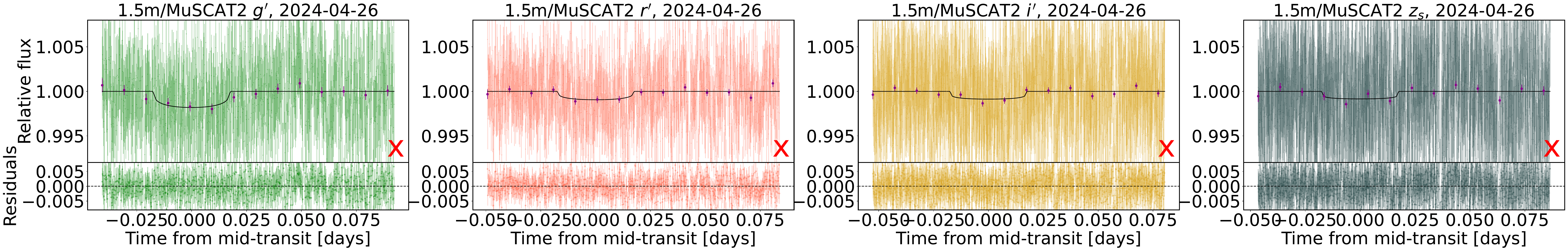}
    \caption{Same as Fig.~\ref{fig:g4529}, but for \plac.}
    \label{fig:g5388}
\end{figure*}

\FloatBarrier

\section{Spectroscopic analysis of \plab}
Additional material regarding the the RV analysis of \plab. 
\begin{table*}[htbp]
    \caption{Comparison of the RV models for \plab.}
    \centering
    \begin{tabular}{l c c c}
        \hline
        \hline
        Model & $\log \mathcal{Z}$ & $3 \sigma \,K$ upper limit (m\,s$^{-1}$) & $\log \mathcal{Z} - \log \mathcal{Z}_{\rm best}$\\
        \hline
        Flat line & $-189.01 \pm 0.05$ & ...  & $-3.37$ \\
        1 planet Keplerian & $-192.03 \pm 0.07$ & $3.35$ & $-6.39$ \\
        1 planet Keplerian + quadratic trend & $-215.77 \pm 0.12$ & $3.25$ & $-30.13$\\
        1 planet Keplerian + Sin for Stellar Activity & $-209.89 \pm 0.46$ & $2.25$ & $-24.21$\\
        1 planet Keplerian + QP Gaussian process & $-185.64 \pm 0.04$ & $2.41$ & $0$ \\
        \hline
    \end{tabular}
    \label{tab:model_comparison}
\end{table*}

\begin{figure*}[h!]
        \centering
        \includegraphics[width=0.9\textwidth]{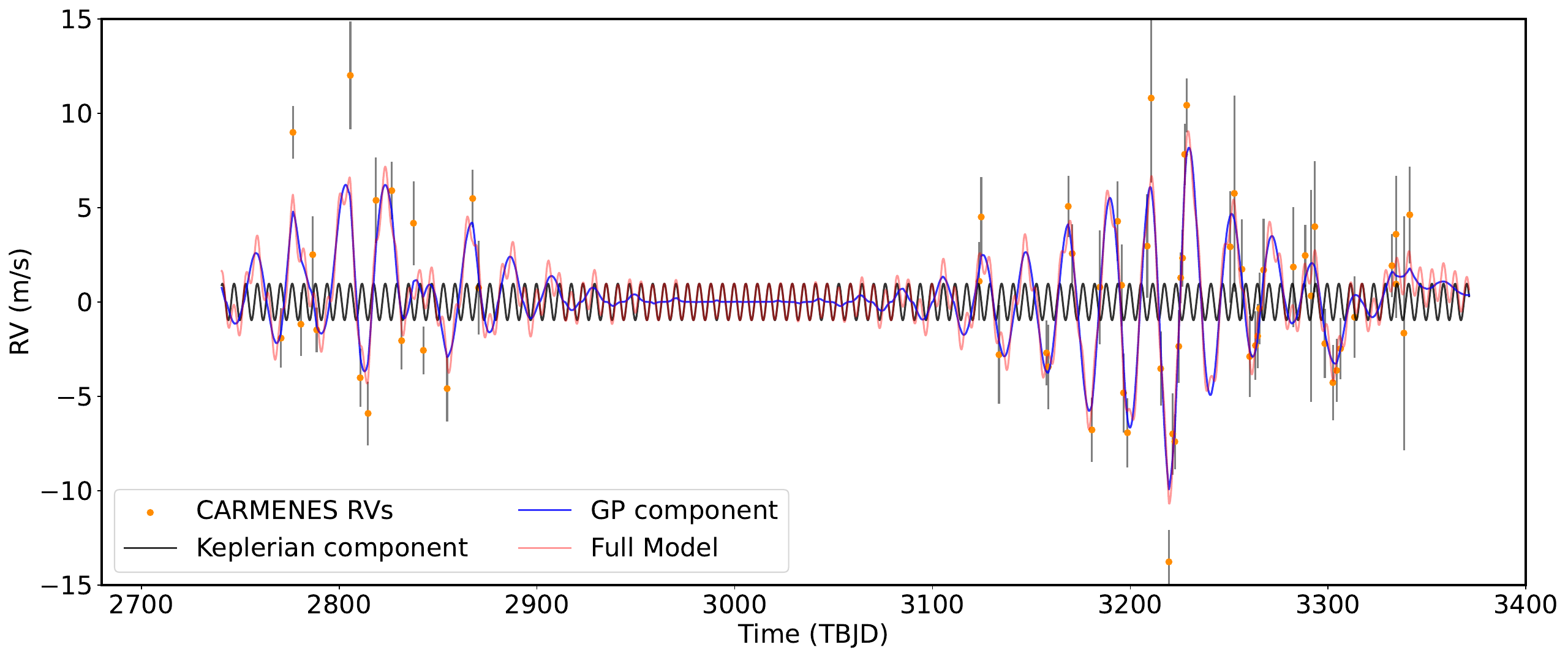}
        \caption{Global model of the RV data for \plab. The blue line shows the GP model fitted using a QP kernel as derived in Sect.~\ref{section:data_analysis}.}
\label{fig:rv_gp_4529}
\end{figure*}

\FloatBarrier
\section{Plots for Atmospheric Modeling}

\begin{figure*}[h!]
\centering
\includegraphics[width=0.49\textwidth]{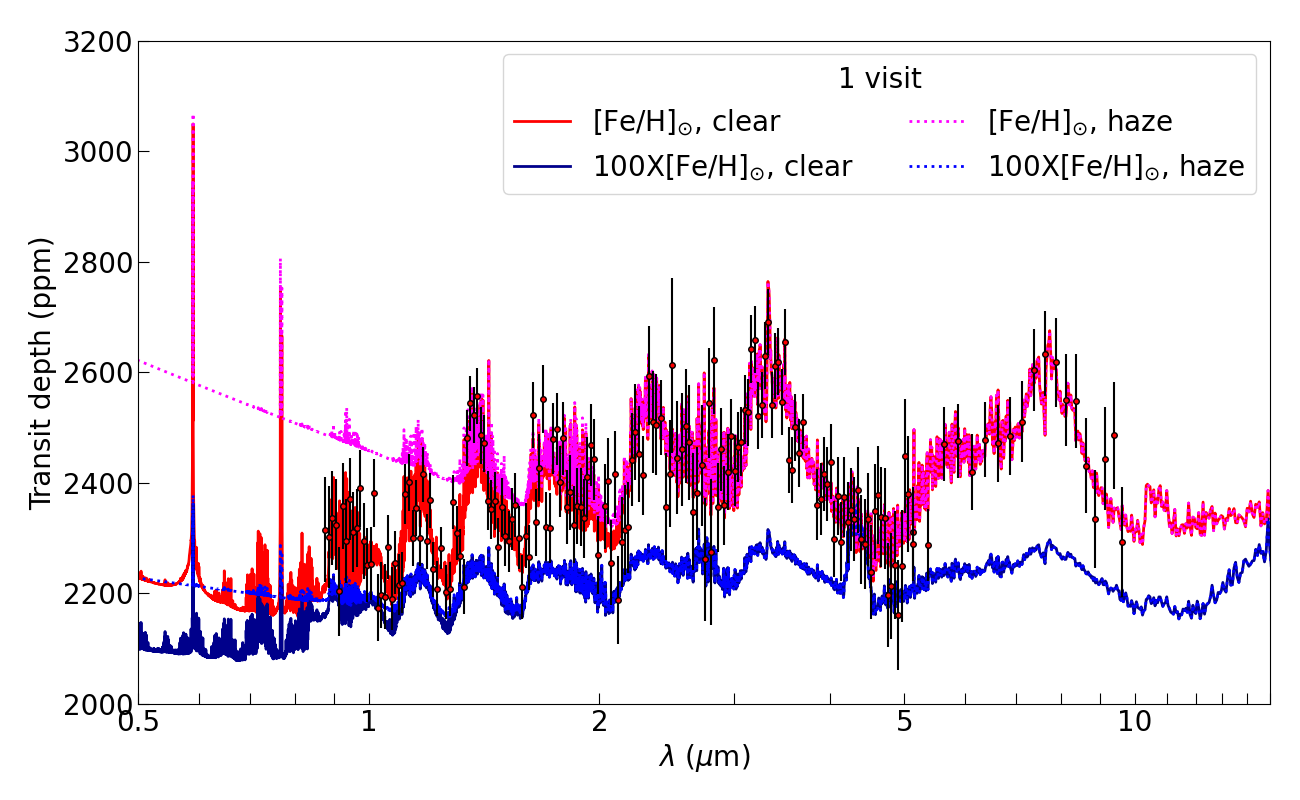}
\includegraphics[width=0.49\textwidth]{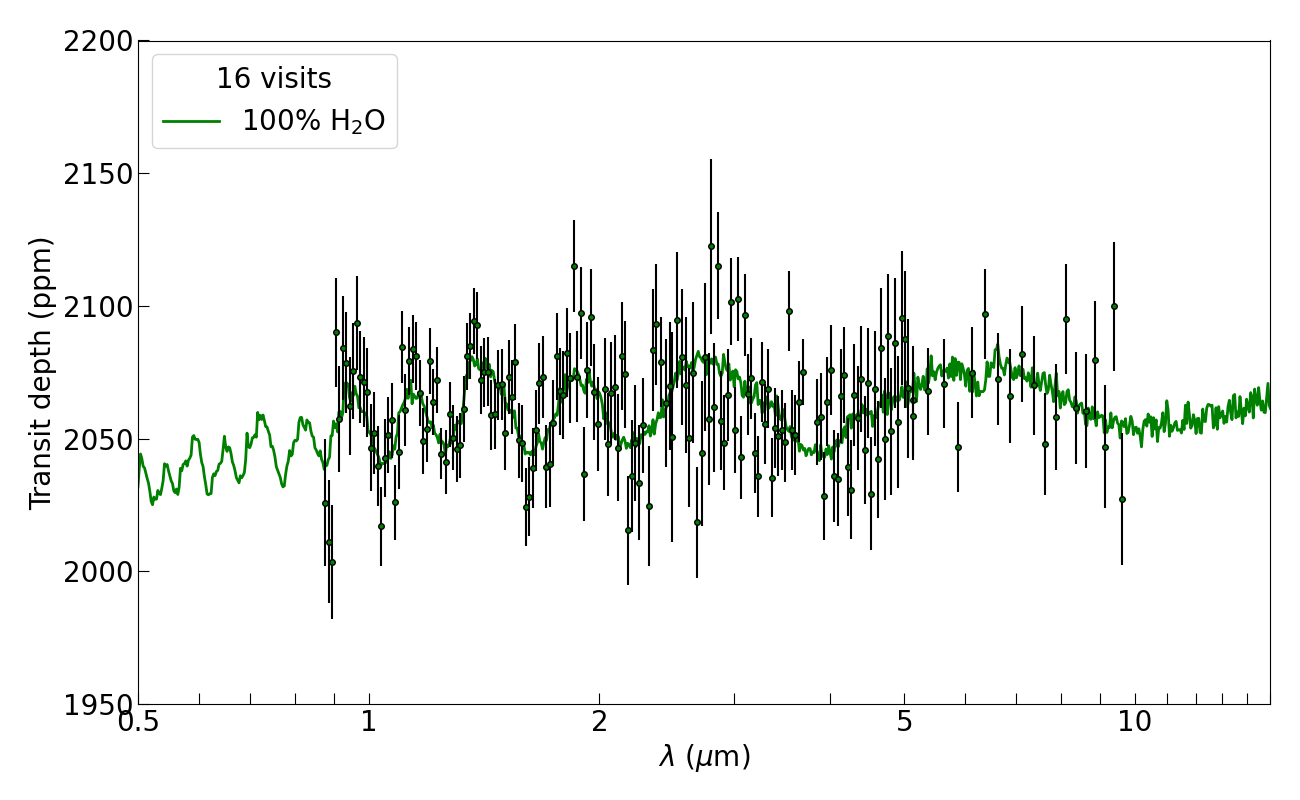}
\caption{Synthetic atmospheric transmission spectra of \plaa. \textit{Left:} fiducial models for clear or hazy H/He atmospheres with scaled solar abundances. \textit{Right:} model for a steam H$_2$O atmosphere. 
Simulated measurements with error bars are shown for the observation of one (\textit{left}) or 16 (\textit{right}) transits with \textit{JWST} NIRISS-SOSS, NIRSpec-G395H, and MIRI-LRS configurations.}
\label{fig:jwst_atmo}
\end{figure*}

We report in Fig.~\ref{fig:jwst_atmo} the results of atmospheric modeling described in Sect.~\ref{ssec:atmospheric_characterization}.

\FloatBarrier

\end{appendix}

\end{document}